\documentclass[preprint,showpacs,letterpaper,preprintnumbers,amsmath,amssymb,nofootinbib]{revtex4}

\usepackage{graphicx}
\usepackage{bm} 
\begin{document}

\title{Gravitational 
magnetic monopoles and Majumdar-Papapetrou stars}

\author{Jos\'e P. S. Lemos}
\email{lemos@fisica.ist.utl.pt} 
\affiliation{Centro Multidisciplinar de Astrof\'{\i}sica - CENTRA, 
Departamento de F\'{\i}sica, Instituto Superior T\'ecnico, Universidade 
T\'ecnica de Lisboa, Av. Rovisco Pais 1,
1049-001 Lisboa, Portugal, \&\\
Observat\'orio Nacional - MCT,\\
Rua General Jos\'e Cristino 77,
20921-400, Rio de Janeiro, Brazil.}
\author{Vilson T. Zanchin\thanks}
\email{zanchin@ccne.ufsm.br}
\affiliation{Departamento de F\'{\i}sica, Universidade Federal de
Santa Maria, 97119-900 Santa Maria, RS, Brazil.}

\begin{abstract}  
During the 1990s a large amount of work was dedicated to studying
general relativity coupled to non-Abelian Yang-Mills type theories.
Several remarkable results were accomplished.  In particular, it was
shown that the magnetic monopole, a solution of the Yang-Mills-Higgs
equations can indeed be coupled to gravitation. For a low Higgs mass
it was found that there are regular monopole solutions, and that for a
sufficiently massive monopole the system develops an extremal magnetic
Reissner-Nordstr\"om quasi-horizon with all the matter fields laying
inside the horizon.  These latter solutions, called quasi-black holes,
although non-singular, are arbitrarily close to having a horizon, and
for an external observer it becomes increasingly difficult distinguish
these from a true black hole as a critical solution is approached.
However, at precisely the critical value the quasi-black hole turns
into a degenerate spacetime.  On the other hand, for a high Higgs
mass, a sufficiently massive monopole develops also a quasi-black
hole, but at a critical value it turns into an extremal true horizon,
now with matter fields showing up outside.  One can also put a small
Schwarzschild black hole inside the magnetic monopole, the
configuration being an example of a non-Abelian black hole.
Surprisingly, Majumdar-Papapetrou systems, Abelian systems constructed
from extremal dust (pressureless matter with equal charge and energy
densities), also show a resembling behavior. Previously, we have
reported that one can find Majumdar-Papapetrou solutions which are
everywhere nonsingular, but can be arbitrarily close of being a black
hole, displaying the same quasi-black hole behavior found in the
gravitational magnetic monopole solutions.  With the aim of better
understanding the similarities between gravitational magnetic
monopoles and Majumdar-Papapetrou systems, here we study a particular
system, namely a system composed of two extremal electrically charged
spherical shells (or stars, generically) in the
Einstein$-$Maxwell$-$Majumdar-Papapetrou theory.  We first review the
gravitational properties of the magnetic monopoles, and then compare
with the gravitational properties of the double extremal electric
shell system. These quasi-black hole solutions can help in the
understanding of true black holes, and can give some insight into the
nature of the entropy of black holes in the form of entanglement.


\end{abstract} 
 
\maketitle

\section{Introduction}

The coupling of general relativity to Yang-Mills $SU(2)$ 
non-Abelian theories
was studied in detail in the 1990s giving rise to a fuller
understanding of the systems involved through a series of remarkable
results.  This effort started after the paper by Bartnick and McKinnon
\cite{bartnickmckinnon}, which showed that Einstein$-$Yang-Mills
theory has one particle solution, the Bartnick-McKinnon particle, in
spite of neither pure gravity nor pure Yang-Mills having a particle
solution on their own.  Further studies inserted a black hole inside
this particle \cite{galtsvo1etal,bizonYMBH} with the conclusion that,
although unstable, the solution could be an instance of no-hair
violation, which in turn motivated new works.  Similar systems were
then studied, such as, the Einstein-Skyrme system
\cite{moss,maedaPRD1,maedaPRD2,moss2,shiiki}, the
Einstein$-$Yang-Mills$-$dilaton system \cite{maedaPRD1}, the
Yang-Mills$-$Proca system \cite{mathur,maedaPRD2}, the
Einstein$-$Yang-Mills$-$Higgs sphaleron system
\cite{mathur,maedaPRD2}, Einstein$-$Yang-Mills in anti-de Sitter 
spacetimes \cite{raduwinstaley}, 
all these systems have in common that their
global electromagnetic type charge is zero, a good review
is in \cite{volkovgaltsov}.
There is yet another very interesting system, which concerns us here,
the magnetic monopole which is a solution of the
Einstein$-$Yang-Mills$-$Higgs system.  Indeed, the 't Hooft-Polyakov
magnetic monopole, is a solution of the pure Yang-Mills$-$Higgs
system (i.e., Einstein$-$Yang-Mills$-$Higgs with zero gravity), when
the Yang-Mills and the Higgs fields are in the adjoint $SO(3)$
representation (see the review paper of Goddard and Olive
\cite{goddardolive}). The 't Hooft-Polyakov magnetic
monopole, at large distances, has the same structure of the Dirac
monopole, however the core is non-singular
\cite{goddardolive,leeweinbergPRL94}. When one couples gravitation, at
least weakly, the magnetic monopole solution is still there, as was
noticed in \cite{newvenhuizenetal76}, now exerting a small
gravitational attraction. For strong gravitational fields the system
was studied much later, in the wake of the Bartnick-McKinnon
particle, notably by Ortiz \cite{ort}, Lee, Nair and Weinberg
\cite{leenairweinbergPRD92,leenairweinbergPRL92}, 
Breitenlohner, Forg\'acs and Maison
\cite{breitenetal1,breitenetal2}, and Aichelburg and Bizon
\cite{bizon}, among others. A distinctive feature of this system is
that it has a global magnetic charge, which of course influences the
properties of the spacetime.  In addition, the
Einstein$-$Yang-Mills$-$Higgs system has two lengthscales, one due to
the mass of the W particle (the Yang-Mills particle  that has eaten 
some mass in the symmetry breaking process), the other due to the mass of the
Higgs. For low Higgs mass, the associated large Compton length scale
does not interfere much, and the structure of the monopole is
characterized in great extent by the W field features. For this
system, the one analyzed in 
\cite{ort,leenairweinbergPRD92,leenairweinbergPRL92,breitenetal1,breitenetal2,bizon}, 
it was found that there are regular solutions, and moreover, for a
sufficiently massive monopole the system turns into an extremal
quasi-black hole, developing an extremal quasi-horizon, with all the
non-trivial matter fields inside it. A quasi-black hole is a
configuration which is non-singular but on the verge of having a
horizon at some radius $r_*$. More specifically, 
quasi-black holes are non-singular solutions 
arbitrarily close to having a horizon. For an external observer 
it becomes increasingly difficult distinguish a quasi-black hole 
from a true black hole as a critical solution is approached. 
At the critical value one has to distinguish two situations. 
In the low Higgs mass situation 
a horizon never forms, 
when the configuration has radius $r_*$ the spacetime is degenerated,  
where the time dimension disappears altogether from a region of
the spacetime. The distinction between a quasi-black hole and a 
true black hole, as well as the 
appearance of a degenerated spacetime, was not clear in the early
works. On the other hand, for high Higgs mass the
system behaves differently as was shown later by Lue and Weinberg
\cite{lueweinberg1,lueweinberg2} (see also the review
(\cite{weinbergreview2001}). In this case, for a sufficiently massive
monopole the system turns into a quasi-black hole, and at 
the critical value, a real extremal magnetic Reissner-Nordstr\"om 
black hole appears, developing a true extremal horizon inside the
monopole core, and moreover, non-Abelian matter fields stick out of
the horizon, in gross violation of the no-hair conjecture.  It was
further found that one could insert a Schwarzschild black hole inside
the monopole without perturbing much its structure, forming a
non-Abelian black hole.  But when the radius of the Schwarzschild
black hole achieved a certain value the horizon would jump into
another extremal quasi-horizon. This happens both in the low Higgs
mass case, as was found by the original authors, as well as in the
high Higgs mass case, as was shown by Brihaye, Hartmann and Kunzin
\cite{brihaye1} where the continuation of the original program has
been carried out.  Other studies connected with magnetic monopoles in
the Einstein$-$Yang-Mills$-$Higgs theory can be mentioned: (i) The
thermodynamical properties of these monopole black holes were further
studied by Maeda et al
\cite{maedaPRL,maedaglobalcharge1,maedaglobalcharge2} in the low Higgs
mass case, and by Lue and Weinberg \cite{lueweinberg2} for high Higgs
mass; (ii) Ridgeway and Weinberg found the existence of
non-spherically symmetric magnetic monopole configurations
\cite{ridgeway}; (iii) Dyonic solutions were found by Brihaye et al
\cite{brihaye2,brihaye3}; (iv) Monopole solutions in other theories
were found, like in a Brans-Dicke theory \cite{maedaPRD99}, and in
$SU(3)$, $SU(5)$, and $SU(N)$ gauge theories
\cite{brihaye4,brihaye5,brihayeN}.

Now, in a different context, the study of the Einstein-Maxwell system
goes back to the origins of general relativity where Reissner in
1916 and Nordstr\"om in 1918 found the Reissner-Nordstr\"om solution
(see \cite{mtw} for the appropriate references), and Weyl studied
axisymmetric gravito-electric vacuum systems in four 
dimensions \cite{weyl}.  A great
development occurred in 1947 when Majumdar \cite{majumdar} and Papapetrou
\cite{papapetrou}, drawing upon Weyl's results, found new four-dimensional 
solutions that represent many particles (from one to infinity), each 
particle with mass equal
to charge, located at any desired position, without spatial symmetry in
the most generic case (see \cite{lemoszanchin} for 
a generalization of Majumdar's \cite{majumdar} 
and Papapetrou's \cite{papapetrou}
works to higher dimensional (${\rm d}>4$) spacetimes).  
The idea is borrowed from Newtonian
gravitation: A particle with mass equal to charge is in equilibrium
with other mass equal to charge particle, and so with many other such 
particles, since the gravitational attraction is balanced by the
electric repulsion. The Majumdar-Papapetrou solutions are the general
relativistic realization of this idea. Now, in Newtonian theory, point
particles are point particles, but in general relativity they can be
black holes. This was clarified by Hartle and Hawking
\cite{hartlehawking} who showed that the vacuum solution represent
extremal Reissner-Nordstr\"om black holes at any spatial position. A
different development was taken by Das \cite{das}, who relying on the
work of Majumdar \cite{majumdar}, put dust particles on the point
particles positions, evading the black hole horizons.  Several other
authors have further analyzed the properties of Majumdar-Papapetrou systems
\cite{cohen,gautreau,guilfoyle,ida,ivanov,varela}. 
Bonnor and collaborators
\cite{bonnor1,bonnor2,bonnor3,bonnor4,bonnornonspherical} in a series
of
papers have shown and studied important new solutions of
Majumdar-Papapetrou equations and properties of the system. 
In particular, in 
\cite{bonnor2,bonnor3,bonnor4}, spherical extremal matter stars 
(where extremal matter stars
are defined as stars composed of matter 
with charge density equal to the energy
density), with an exterior extremal Reissner-Nordstr\"om metric, were
found. These are the Bonnor stars. Bonnor stars were further developed 
by Lemos and Weinberg \cite{lemosweinberg} where new explicit 
solutions were found. It was also found that these new stars, 
as well as Bonnor stars, develop 
a quasi-black hole behavior, and there are cases that the 
solution can even display some kind of hair \cite{lemosweinberg}.  In
addition, in \cite{kleberlemoszanchin} a thick shell solution was
found. In the limit of zero interior
radius for this thick shell, the solution is a Bonnor star, in the
limit of the thickness going to zero, the solution is a thin shell. 
These solutions also have quasi-black hole behavior.

Here we want to explore further the analogy between gravitational
magnetic monopoles and Majumdar-Papapetrou stars.  In the previous
papers \cite{lemosweinberg,kleberlemoszanchin}, the
Majumdar-Papapetrou solutions found, although complex, did not exhibit
the full behavior of the gravitational magnetic monopoles, where there
is an interplay between the W-field scale and the Higgs field scale.
We construct here a Majumdar-Papapetrou system which shows such a full
behavior.  Such a system is composed of two infinitesimally thin
shells.  Majumdar-Papapetrou thin shells have many interesting
properties. Let us think first of one thin shell to simplify. We will
call it the star, it is a regular solution.  Fix the mass of the star,
and study the set of formed configurations as one decreases its
radius. For a sufficiently small radius the star develops an extremal
Reissner-Nordstr\"om quasi-black hole.  The same happens if instead
one fixes the radius and increases the mass.  One can go further and
put another thin shell inside the thin shell star.  One can then ask,
when the radii of the system are decreased which shell is going to form
a quasi-horizon first?  The usual case is the outer shell developing a
quasi-horizon first, the whole system being inside the quasi-black
hole.  But, depending on the parameters and constraints, the inner
shell can develop a quasi-horizon first, in which case we have an
extremal quasi-black hole in the core of the system, with star matter
floating outside.  One can also put an extremal Reissner-Nordstr\"om
true black hole inside the regular star (as was done in the
gravitational magnetic monopole case, when one puts a small
Schwarzschild black hole inside the magnetic monopole) and then
increase the black hole radius through a set of configurations.  At a
certain point the whole system jumps into a new extremal
Reissner-Nordstr\"om quasi-black hole.
If we exchange star for monopole, the properties of this Majumdar-Papapetrou
system are identical to the properties of the gravitational magnetic
monopole system.  All these similarities with the gravitational
magnetic monopole will be explored in this paper.
A similarity which we do not explore, 
is that both systems permit non-spherically symmetric solutions, 
in the magnetic monopole case see \cite{ridgeway}, 
in the Majumdar-Papapetrou case see \cite{bonnornonspherical}. 
We note that the Majumdar-Papapetrou solutions, such as the 
extremal Reissner-Nordstr\"om black hole solutions and the Bonnor stars, 
are also of interest in extensions of general relativity, 
since the system turns out to be supersymmetric 
when embedded in a larger theory, such as $N=2$ gauged supergravity 
(see \cite{peet} for a review of Majumdar-Papapetrou solutions in 
supergravity and string theories).

The paper is organized as follows. In section II we overview 
the properties of gravitational magnetic monopoles that 
most interest us, we give the equations and define the 
important scales, we review the low Higss field (low $b$) case without, 
and then with, an interior Schwarzschild black hole, and review also 
the high $b$ case. In section III we study the properties 
of the Majumdar-Papapetrou two shell system: we give the 
equations and length scales, assume some constraints for 
the shells and present the solution, study the 
equivalent low $b$ behavior without and with an interior 
extremal Reissner-Nordstr\"om true black hole, 
and then the equivalent high $b$ behavior. A remark: when we write a 
black hole it means a true extremal Reissner-Nordst\"om black hole, 
when we write a quasi-black hole it means solutions of matter 
configurations that are on the verge 
of being a black hole. In some instances, quasi-black holes turn into 
degenerated spacetimes \cite{leenairweinbergPRD92,lemosweinberg},  
in other instances turn into real black holes \cite{lueweinberg2}.

\section{Gravitational behavior of magnetic monopoles, an 
overview} 

In this section we overview the solutions for gravitational magnetic
monopoles.  The logical presentation of the material reflects in a
unified way the work of the authors on this subject and is suited for
comparison with the subsequent analysis on Majumdar-Papapetrou stars.

\subsection{The Einstein$-$Yang-Mills$-$Higgs magnetic sector}

\subsubsection{The action and equations of motion}

The action of the Einstein$-$Yang-Mills$-$Higgs  
theory is  $(G=c=1)$ 
\begin{equation}
S=\int d^4x\sqrt{-g}\left(-\frac{1}{16\,\pi}\,R+
{\cal L}_{\rm matter}\right)
\,,
\label{actionEYMH}
\end{equation}
where $R$ is the scalar curvature, and ${\cal L}_{\rm matter}$ 
is the Yang-Mills$-$Higgs Lagrangian given by
\begin{equation}
{\cal L}_{\rm matter}=-\frac{1}{4}F_{\mu\nu}^aF^{a\,\mu\nu}+
\frac12\,D_\mu\phi^a\,D^\mu\,\phi^a-\frac{\lambda}{2}
\left({\phi^a}^2-v^2\right)^2
\,,
\label{L-EYMH}
\end{equation}
\begin{equation}
F_{\mu\nu}^a=\partial_\mu A_\nu^a-\partial_\nu  A_\mu^a -
e\, {\epsilon^{a}}_{bc} A_\mu^bA_\nu^c
\,,
\label{F-equation}
\end{equation}
\begin{equation}
D_\mu\phi^a=\partial_\mu\phi^a-e\,{\epsilon^{a}}_{bc}A_\mu^b\phi^c
\,,
\label{phi-equation}
\end{equation}
where $e$ is the gauge coupling constant, $\lambda$ the Higgs
coupling constant, and $v$ the vacuum expectation value of the 
Higgs field. The Yang-Mills connection $A^a$ and the Higgs field 
$\phi^a$ take values on the Lie algebra of the $SU(2)$ group, 
with ${}^a$ being an internal index. 
The potential 
$\frac{\lambda}{2}\left({\phi^a}^2-v^2\right)^2$ 
in the matter Lagrangian 
has a family of gauge-equivalent minimums, given by ${\phi^a}^2=v^2$, 
which breaks spontaneously the $SU(2)$ symmetry down 
to $U(1)$. One can choose the vacuum to be in 
the third internal direction $\phi^a=v\,\delta^{a3}$ 
(for details see \cite{goddardolive,weinbergreview2001}). 
The elementary particles of the theory 
are  the electromagnetic $U(1)$ massless gauge field
(a photon), two massive W particles with charge $\pm e$ 
and mass $m_{\rm W}=e\,v$, and the neutral massive field 
$\phi^3$ with mass $m_{\rm H}=\frac{1}{\sqrt\lambda\,v}$.
There is also the massless graviton.

The monopole configuration is spherically symmetric 
with metric written generically in terms of two functions 
$A(r)$ and $B(r)$ as 
\begin{equation}
ds^2=-B(r)\,dt^2+A(r)\,dr^2+
r^2\left(d\theta^2+\sin^2\theta\,d\phi^2\right)
\,,
\label{metricEYMHSch}
\end{equation}
with a magnetic Yang-Mills field, written in terms of 
one function $u(r)$, as 
\begin{equation}
A_0=0\;\;,\quad
A_i^a=\epsilon_{iaj}\,{\hat r}^j\frac{1-u(r)}{e\,r}\,,
\label{YMfieldEYMHspherical}
\end{equation}
and a Higgs field, written in terms of one function $h(r)$, as
\begin{equation}
\phi^a=v\,{\hat r}^a h(r)\,,
\label{higgsfieldEYMHspherical}
\end{equation}
where $\epsilon_{iaj}$ is the Levi-Civita tensor, and 
${\hat r}$ is the unit vector in the radial direction. 
Putting this ansatz into the 
Einstein$-$Yang-Mills$-$Higgs action and 
varying the action with relation to the 
four functions yields 
four equations, two for the gravitational 
fields $B(r)$ and $A(r)$, one for the 
Yang-Mills field 
$u(r)$, and one for the Higgs field $h(r)$. 
The equations are, respectively, (see \cite{leenairweinbergPRD92}),
\begin{equation}
\frac{(AB)'}{AB}=16\pi\,r\,
\left(\frac{u'\,^2}{e^2r^2}+
\frac12\,v^2\,h'\,^2\right)
\,,
\label{equationforB}
\end{equation}
\begin{equation}
 \left[r\left(1-\frac{1}{A}\right)\right]'
= 8\pi\,r^2\left[\frac{1}{A}
\left(\frac{u'\,^2}{e^2r^2}+
\frac12\,v^2\,h'\,^2\right)
+\frac{(u^2-1)^2}{2\,e^2\,r^4}+
\frac{u^2\,h^2\,v^2}{r^2}+
\frac{\lambda}{2}v^2(h^2-1)^2\right],
\label{equationforA}
\end{equation}
\begin{equation}
\frac{1}{\sqrt{AB}}\left[\frac{\sqrt{AB}\,u'}{A}\right]'
=\frac{u\,(u^2-1)}{r^2}+e^2u\,h^2\,v^2
\,,
\label{equationforu}
\end{equation}
\begin{equation}
\frac{1}{r^2\sqrt{AB}}
\left[\frac{r^2\sqrt{AB}\,h'}{A}\right]'
=\frac{2hu^2}{r^2}+2\,\lambda\,h(h^2-1)\,v^2
\,.
\label{equationforh}
\end{equation}
Sometimes, instead of $A(r)$ 
it is used the mass function $m(r)=r\left(1-\frac{1}{A(r)}\right)$.
There are four parameters in the theory: $G$ (which we have set 
equal to one), $e$, $\lambda$ and $v$. With these parameters 
one can form two dimensionless parameters, $\alpha$ and $\beta$. 
Since $v$ is dimensionless, it is already 
a sought parameter, 
$\alpha=v$. The other dimensionless parameter is 
$\beta=\frac{\sqrt{\lambda}}{e}$.
(In passing, note that in these studies 
of gravitational magnetic monopoles, $G$ is not 
usually set to one, but rather 
$G=m_{p}^{-2}=l_{p}^{2}$ $\,(\hbar=c=1)\,$, and  
$m_{p}$ and $l_{p}$ are the Planck mass and 
the Planck length, respectively. Here we are putting 
$m_{p}=1$ and $l_{p}=1$, i.e., we are measuring 
everything in terms of these scales. It is straightforward 
to move from one system of units to the other: 
Every time one finds a mass one should divide by $m_p$, 
every time one finds a length one should divide by $l_p$,
in the end collect all $m_p$s and $l_p$s, 
transform  $l_p$ into $m_p$ by $l_p=m_p^{-1}$, 
and put back $G$ from $m_p^{-2}=G$).

\subsubsection{Some properties and scales of the magnetic 
monopole}

The magnetic monopole solution can be 
characterized by its mass and radius, and by a secondary 
mass and a secondary radius. 
To understand the effects of gravity it is useful to rewrite 
the parameters $\alpha$ and $\beta$, defined above, 
in terms of two renewed parameters, $a$ and $b$, defined through the
characteristic masses and radii themselves. 
Indeed, for a weak gravitational field the 
magnetic charge of the monopole is $Q_{\rm m}=\frac{1}{e}$, 
its radius is given roughly by the Compton wavelength of the Yang-Mills field, 
$r_{\rm m}\simeq \frac{1}{e\,v}$, and its mass by 
the magnetic energy 
$M_{\rm m}\simeq \frac{Q_{\rm m}^2}{r_{\rm m}}=\frac{v}{e}$. 
Thus, instead of $\alpha$ given above ($\alpha=v$),
we can define a parameter $a$ (with $a\sim v^2$) as 
\begin{equation}
a\equiv \frac{M_{\rm m}}{r_{\rm m}}
\,,
\label{definitionofa}
\end{equation}
which is a useful characterization when we turn on gravitation, and 
for later comparison. The other parameter $b$ can also 
be written in similar terms: Since there is the Higgs mass 
scale, the monopole solution has secondary mass and radius scales. 
The secondary radius is given by the 
Compton wavelength of the Higgs field, 
$r_{\rm m_2}\simeq\frac{1}{\sqrt\lambda\,v}$, and the secondary mass 
is given by 
$M_{\rm m_2}\simeq \frac{Q_{\rm m}^2}{r_{\rm m_2}}=
\frac{\sqrt\lambda\,v}{e^2}$. Thus, instead of $\beta$  
given above ($\beta=\frac{\sqrt\lambda}{e}$),
we can define a parameter $b$ (with $b\sim\frac{\lambda}{e^2}$) 
as
\begin{equation}
b \equiv\frac{M_{\rm m_2}/r_{\rm m_2}}{M_{\rm m}/r_{\rm m}}
\,,
\label{definitionofb}
\end{equation}
which displays the coupling between both, mass over radius scales. 
There are  three parameters $v,\,e,\,\lambda$ and four quantities, 
$M_{\rm m},\,r_{\rm m},\,M_{\rm m_2},\,r_{\rm m_2}$, 
so there is an equation 
\begin{equation}
M_{\rm m}\,r_{\rm m}=M_{\rm m_2}\,r_{\rm m_2}
\,,
\label{constraint1}
\end{equation}
which constrains the four quantities. For instance, $r_{\rm m_2}$ can be 
considered as fixed once the other three quantities are known.

\subsection{The gravitational behavior as a function of $a$ (gravitation) 
for low $b$ (low Higgs mass)}

Here, we overview the solutions found in
\cite{ort,leenairweinbergPRD92,leenairweinbergPRL92,breitenetal1,breitenetal2}
keeping in mind that we will later want them for comparison.  We
present the metric and matter functions as a function of radius,
discuss the naked horizon behavior and the Coulumb character of this
type of solutions, put a Schwarzschild black hole inside, and resketch
some diagrams covering the space of solutions. 

From the last subsection, low $b$ indicates a small Higgs mass $m_{\rm
H}$, or large associated Compton wavelength, which means that the
Higgs does not participate in the dynamics, it has very little
influence on the monopole structure. Reinterpreted through Equation
(\ref{definitionofb}) one can also see the low $b$ case as a monopole
with small secondary mass $M_{\rm m_2}$ or large secondary radius
$r_{\rm m_2}$. Given a low $b$ configuration, we want now to understand
how the structure changes as gravity is turned on higher and higher,
i.e., as the parameter $a=\frac{M_{\rm m}}{r_{\rm m}}$ increases.  

\subsubsection{The regular magnetic monopole solution: from no gravitation 
to the extremal quasi-black hole}

\begin{figure} [t]
\includegraphics*{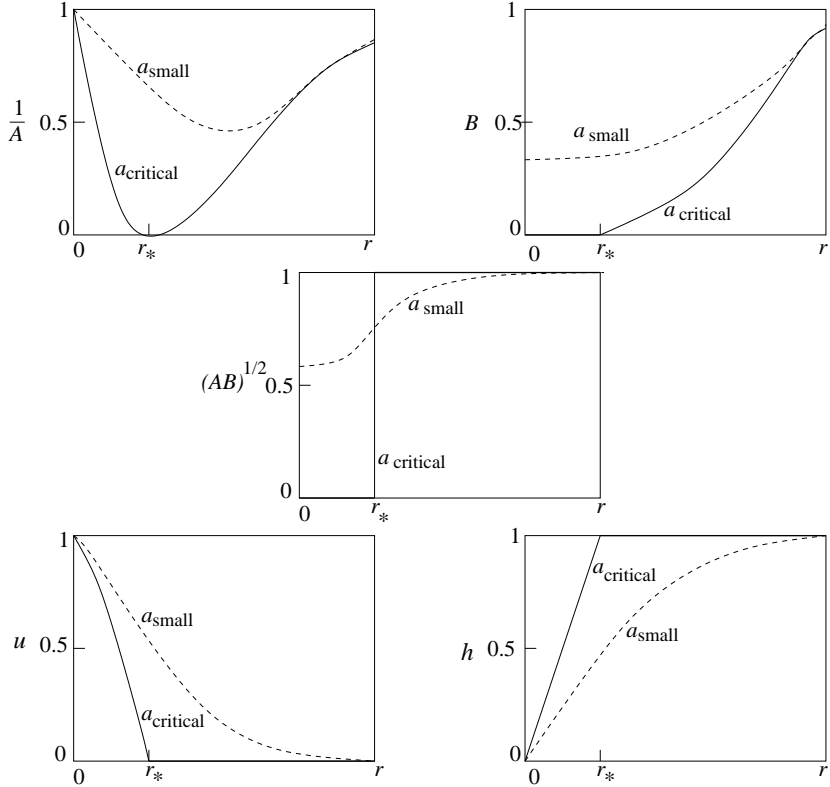}
\caption{\label{lowbmonopole}
The graphs of the metric and matter functions, 
$(1/A,B,(AB)^{1/2},u,h)$, 
are plotted as a function of $r$ in the low $b$ case. The curves $a$ small 
are typical of small gravitational effects, and the curves $a$ critical 
give the properties of the extremal quasi-black hole. The radius $r_*$ is 
the the radius at which the quasi-horizon is formed,
see text for details.}
\end{figure}

Let us start with $a$ (see Equation (\ref{definitionofa})) 
small. This means a highly dispersed magnetic 
monopole with small mass $M_{\rm m}$ and large radius $r_{\rm m}$. As
$a$ increases the solution gets more general relativistic and
eventually should get to a black hole, where $a=a_{\rm crit}\sim1$ (for
instance, if the configuration formed a Schwarzschild black hole
$a_{\rm crit}=\frac12$, or if it formed an extremal Reissner-Nordstr\"om
black hole $a_{\rm crit}=1$). It does not happen exactly like this. 
The solution in
the limit of $a_{\rm crit}$ yields a quasi-black hole as defined in
\cite{lueweinberg2,lemosweinberg}.  To get a grip on the solutions we
draw in Figure \ref{lowbmonopole} diagrams showing the metric functions
and the matter field functions as a function of $r$ for two values of
$a$, $a$ small and $a_{\rm crit}$ \cite{leenairweinbergPRD92}.  The
function $A$ signals the existence of a black hole horizon, 
the function $B$ is the redshift function, 
the product function $(AB)^{1/2}$ tells whether a horizon
is naked or not, and the functions $u$ and $h$ report on the hair or
no-hair of the solution. More specifically:
(i) The function $A$, or better $1/A$, indicates how strong the
curvature is, and in particular indicates the existence of a black hole
horizon. At $r=0$, $1/A$ should be 1 in order that there are no conical
singularities, and at $r\rightarrow\infty$ should be again 1 for
asymptotically flat spacetimes. Now, for $a\simeq0$ spacetime is flat
and $1/A\simeq1$ for all $r$. For small $a$ there is a small dip at
intermediate $r$ as shown in Figure \ref{lowbmonopole}. For large $a$
the dip is large, and for $a=a_{\rm crit}$, $1/A$ is zero, indicating  
that a black hole horizon might have formed, here an extremal one
since $1/A$ gets a double zero. In fact, 
as was first noticed in \cite{lueweinberg2}, a true extremal 
black never forms.  Instead, for a configuration with a radius 
arbitrarily near the critical radius, 
a quasi-black hole forms (i.e., a matter  
solution whose gravitational properties are virtual indistinguishable 
from a black hole \cite{lueweinberg2}), and at $a_{\rm crit}$ precisely, 
a degenerate spacetime appears as it is found when one looks to the 
metric function $B$. 
(ii) The metric function $B$ gives the redshift behavior, or the
relative behavior of clocks at different spatial positions. It is the
function that distinguishes a true black hole from a quasi-black hole,
as we will now see.  For $a\simeq0$,  one has $B\simeq1$. For small $a$, $B$
lowers at the origin showing the existence of a gravitational
potential, and goes to 1 at $r\rightarrow\infty$. For $a=a_{\rm crit}$
or very near it, $B$ goes again to one at $r\rightarrow\infty$, but
now it is zero up to the monopole radius. This is odd, the infinite
redshift surface is not a surface it is a three-dimensional region.
To be a black hole $B$ should go to zero at a given $r$ only.  This
means that the solution at $a=a_{\rm crit}$ does not represent a
smooth manifold. Thus, the quasi-black hole configuration gives rise 
to a degenerated spacetime.  For $a$ very near the critical
value it is very hard to distinguish the quasi-black hole solution
from a true black hole.  The radius of these quasi-black holes is
denoted by $r=r_*$, and is arbitrarily near to the radius of the
extremal Reissner-Nordstr\"om black hole of same mass and charge, see
Figure  \ref{lowbmonopole}.
(iii) One should also pay attention to the behavior of $(AB)^{1/2}$,
which says whether the horizon is naked or not, as it will be precised
below.  For $a\simeq0$ one has $(AB)^{1/2}\simeq1$. For $a$ small, 
$(AB)^{1/2}$ is
small at $r=0$ and 1 at $r\rightarrow\infty$. For $a=a_{\rm crit}$, 
$(AB)^{1/2}$ 
is $0$ up to $r=r_*$ and then steps into $1$, see Figure  
\ref{lowbmonopole}.  
It is interesting to comment further on the behavior of $(AB)^{1/2}$
and its consequences. For the Schwarzschild and Reissner-Nordstr\"om
solutions $(AB)^{1/2}$ is 1 for all radii. However, it is not so here,
as can be seen directly in Figure \ref{lowbmonopole}. The fact that
$(AB)^{1/2}\rightarrow0$ for $r\leq r_*$ at the critical solution
implies that the black hole horizon formed has a naked behavior
\cite{lueweinberg2}.  This means that the components of the Riemann
tensor at the horizon in an orthonormal frame blow up at the
horizon. It can be understood as follows. Suppose a particle sent in
through the monopole, by a distant observer, turns around, and comes
back to the point where it started.  Suppose also the monopole is on
the verge of forming a horizon, i.e., the monopole surface is a
quasi-horizon.  Due to the very small value of $(AB)^{1/2}$ inside and
at the quasi-horizon (see Figure \ref{lowbmonopole}), one finds that
the proper time the particle takes for the round trip is given by
$\Delta \tau\sim r_*\,\epsilon^{q}$, where $r_*$ is the radius of the
quasi-horizon, $\epsilon\equiv (1/A)_{\rm min}$ is a very small
quantity near the critical solution, and $q$ is found by numerical
methods to be $\sim 0.7-1.0$ \cite{lueweinberg2}.  So the particle
takes virtual zero time within the quasi-horizon.  This fact is
related to the black hole nakedeness.  The Riemann tensor on a
particle gives essentially the tidal forces in the particle. It can be
shown that the Riemann tensor in these cases is inversely proportional
to the square of the proper time it takes the particle to cross the
region \cite{lueweinberg2}. Thus if the proper time is zero, the
Riemann tensor, and thus the tidal forces are huge, giving rise to a
naked behavior, the horizon is exposed.  Here $R_{\hat t\hat i\hat
t\hat i}\sim \epsilon^{-2q}$, where $\,\hat{}\,$ means calculated in
the freely falling frame, and the indexes $i$ are spatial indexes. 
So, these are naked black holes. 
Note that for the Schwarzschild,
Reissner-Nordstr\"om, and extremal Reissner-Nordstr\"om black holes,
the Riemann tensor calculated in 
the frame falling with the particle is well behaved, so the horizon is
well behaved, a result that is known otherwise.  The other interesting
time to compute in the round trip is the coordinate time. It is given
by $\Delta t\sim r_*\,\epsilon^{-q}$. Thus, for a coordinate observer, 
the particle takes a long
time to return.  This coordinate time can be
important for entropic considerations \cite{lueweinberg2}. 
(iv) The function $u$ for the Yang-Mills field shows for $a$ small 
a $1/r^2$ fall off for large $r$, and for $a=a_{\rm crit}$ it disappears 
for radii grater than $r=r_*$. (v) The function $h$ 
of the Higgs field for small $a$ is zero and then grows to 
pick up the Higgs vacuum value at large $r$. For $a=a_{\rm crit}$ 
it grows from $0$ at the origin to $1$ at the horizon, 
and stays at 1 up to infinity. This means that there is no hair, 
outside the horizon, only the trivial magnetic and 
vacuum Higgs fields. These quasi-black holes have been 
termed Coulumb type quasi-black holes since they show a 
Coulumb (no hair) field when they form \cite{lueweinberg1}.

\subsubsection{Non-regular magnetic monopoles: 
The Schwarzschild black hole solution inside the monopole}

Up to now we have mentioned the behavior of regular gravitating
monopoles, i.e., solutions that are regular from the origin to
infinity.  One can now put a small
Schwarzschild black hole, with mass $M_{\rm bh}$ and radius $r_{\rm
bh}$, inside the magnetic monopole.  This system is an example of a 
non-Abelian black hole with hair.
One could think that putting a
Schwarzschild black hole inside the monopole would disrupt the
structure, and turn the monopole solution into a time-dependent one
with the Yang-Mills and Higgs fields being accreted onto the black
hole.  But this is not the case, matter, with energy density $\rho$
and radial pressure component $p_{rr}$, can coexist with an event
horizon at its location as long as $\rho+p_{rr}=0$, a result that
follows directly from the conservation equation
${T^{\mu\nu}}_{;\nu}=0$.  A well known example is the Schwarzschild-de
Sitter solution, where the cosmological constant term $\Lambda$ can be
seen as a fluid which certainly obeys $\rho+p_{rr}=0$.  Following 
\cite{leenairweinbergPRD92,maedaPRL} one finds that the non-Abelian
structure inside the monopole may be approximated as a uniform vacuum
energy density $\rho_{\rm vac}$ up to the monopole radius $r_{\rm m} $
such that the black hole in this region has a metric identical to the
Schwarzschild-de Sitter black hole.  For small black holes the
condition  $\rho+p_{rr}=0$ is obeyed and they can 
inhabit the center of the monopole,
i.e., small black holes inside do not perturb much the solution. 
However, when the Schwarzschild black
hole is large enough, such that its mass is of the order of the mass
of the system, the system itself collapses giving rise to a magnetically
charged extremal Reissner-Nordstr\"om quasi-black hole. 
We note that the literature is not clear whether it forms a 
true extremal black hole or an extremal 
quasi-black hole, however by continuity from 
the regular case one is entitled to infer that it is a quasi-black hole, 
followed by a degenerated spacetime at the critical value. 
The appearance of this quasi-black hole happens for a
critical value of the parameter $a$, with 
$a_{\rm crit}\sim1$, or alternatively, 
for a critical value of the total mass $M$, with $M=M_{\rm
m}+M_{\rm bh}$.  The behavior is thus analogous to the regular
monopole in the sense that as one increases gravitation, i.e., as the
parameter $a$ or the mass $M$ of the system increases, one finds a
a magnetically charged extremal Reissner-Nordstr\"om
quasi-black hole.

\begin{figure} [t]
\includegraphics*[height=2.5in]{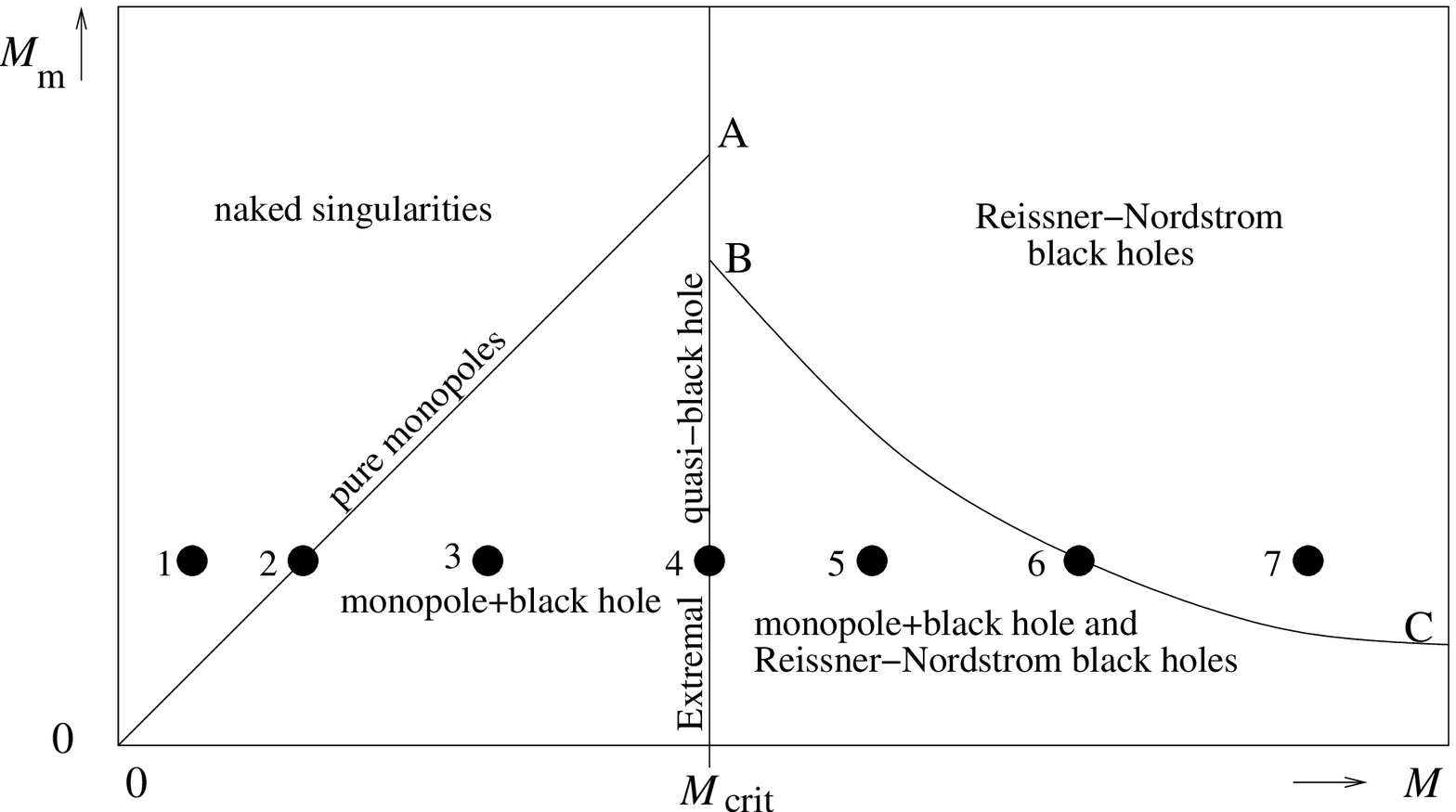}
\vskip 5mm
\includegraphics*[height=1.5in]{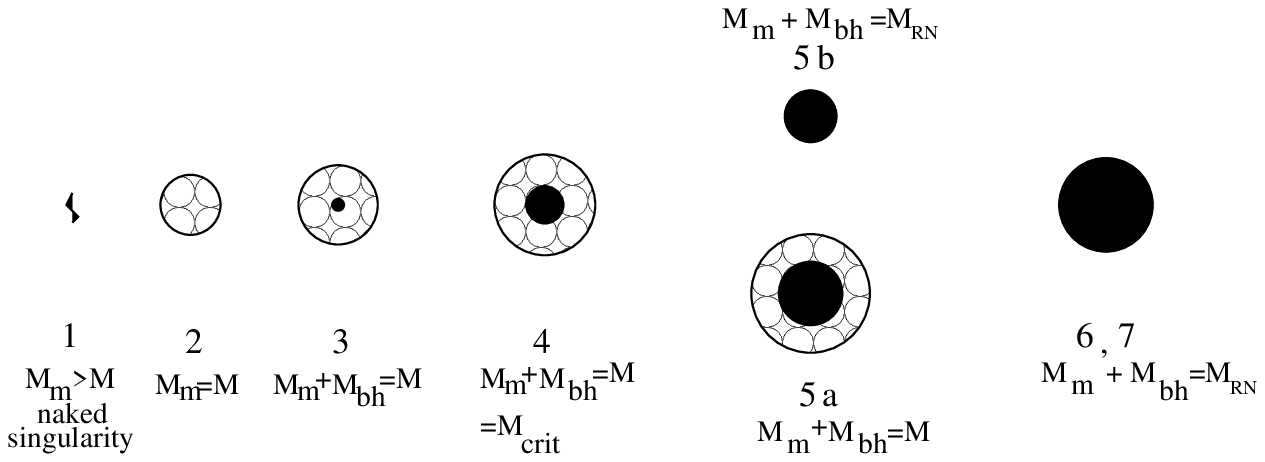}
\caption{\label{graphmonopolemassversustotalmass}
The space of solutions in a $M_{\rm m}\times M$, where 
$M=M_{\rm m}+M_{\rm bh}$ is the total mass, is plotted 
(see also \cite{leenairweinbergPRD92}). 
For each point 1-7, along a constant
monopole mass, in the diagram, the corresponding 
configuration is pictorially represented in the bottom part 
of the figure.}
\end{figure}
\begin{figure} [t]
\includegraphics*[height=2.5in]{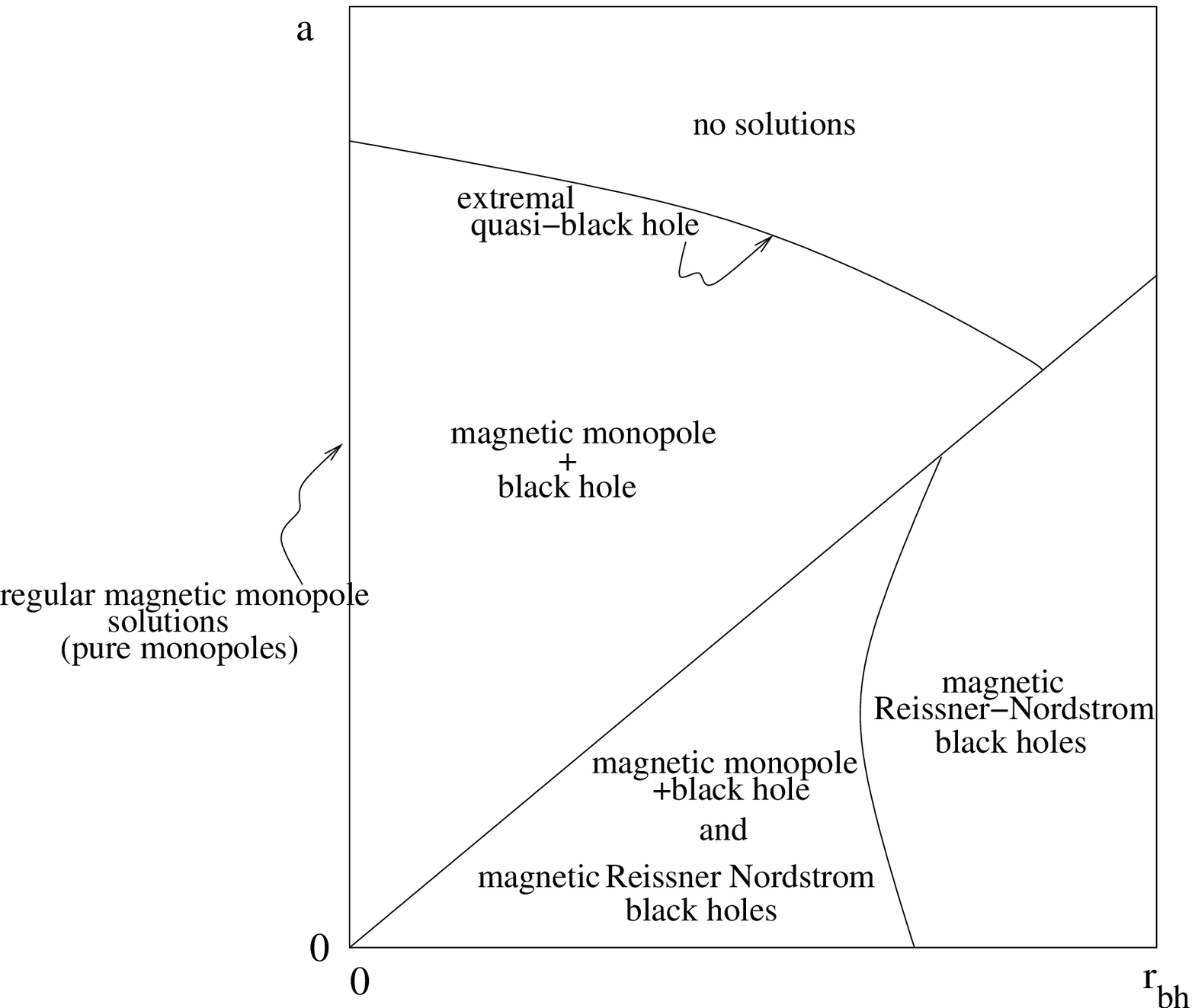}
\caption{\label{aversusrblackhole}
The space of solutions in a $a\times r_{\rm bh}$
diagram (see also \cite{breitenetal1}).}
\end{figure}

To understand the generic behavior it is helpful to make a plot of the
solution space.  One such plot is given in a $M_{\rm m}\times M$,
where $M=M_{\rm m}+M_{\rm bh}$ is the total mass. This is shown in
Figure \ref{graphmonopolemassversustotalmass}, see also
\cite{leenairweinbergPRD92}. There are four areas and three lines.
The pure monopole line, (the regular solutions discussed above) with
the total mass equal to the monopole mass, is represented by a line
with slope 1.  The top-left region represents naked singularities.
The center-left area represents monopole+Schwarzschild (non-Abelian)
black hole solutions mentioned above. At arbitrarily near 
the critical mass the
solutions are extremal magnetic charged Reissner-Nordstr\"om
quasi-black holes (and at precisely the critical mass they turn into 
degenerated solutions).  To the right there is a region where
monopole+Schwarzschild (non-Abelian) black holes coexist with magnetic
(Abelian) Reissner-Nordstr\"om black holes. Then to the far right and
above there is a region of magnetic (Abelian) Reissner-Nordstr\"om
black holes alone.  In this diagram, solutions with a constant black
hole mass are represented by lines parallel to the pure monopole line,
i.e., lines of slope 1.  Lines of constant monopole mass are
horizontal lines.  We show pictorially each representative
configuration along a constant monopole mass line. Each numbered point
(from 1 to 7) in Figure \ref{graphmonopolemassversustotalmass} is
represented in the bottom of the figure by a schematic drawing. In this
drawing, note that the horizon area of the solution containing a
Schwarzschild black hole surrounded by monopole matter (numbered 5a)
is larger than the horizon area of the pure magnetic
Reissner-Nordstr\"om horizon (numbered 5b).  Following the area law,
the smaller one is prone to be unstable and decay to the larger hairy
one. This has interesting implications in the ultimate fate of the
black hole through Hawking evaporation \cite{leenairweinbergPRL92}.

Another similar but interesting plot is $a\times r_{\rm bh}$ diagram,
shown in Figure \ref{aversusrblackhole}, see also \cite{breitenetal1}.
There are four areas and four lines.  There is the pure monopole line
($r_{\rm bh}=0$), which yields the regular solutions discussed above.
The top-left area is the region of no solutions.  
There is the center-left area of
monopole+Schwarzschild (non-Abelian) black hole solutions 
discussed above, there is the bottom-left area where one finds
monopole+Schwarzschild (or non-Abelian) black hole solutions as well
as magnetic (Abelian) Reissner-Nordstr\"om black holes, and then the
right area of magnetic (Abelian) Reissner-Nordstr\"om black holes. The
other lines are boundaries between these areas.

\subsection{The gravitational behavior as a function of $a$ (gravitation) 
for high $b$ (high Higgs mass)}

High Higgs mass reserves surprises.  Here we overview the solutions
found in \cite{lueweinberg1,lueweinberg2} still keeping in mind that we
will later need them for comparison.  High $b$ means
$b\buildrel>\over\sim 40$ \cite{lueweinberg1}. 
From subsection IIA, high $b$ indicates a large Higgs mass $m_{\rm H}$,
or small associated Compton wavelength. This means that the Higgs field
does participate in the dynamics, and can have great influence on the
monopole structure.  Reinterpreted through Equation
(\ref{definitionofb}) one can also see high $b$ as a monopole with
large secondary mass $M_{\rm m_2}$, or small secondary radius $r_{\rm
m_2}$.  In order to understand how the structure changes as gravity is
turned on higher and higher one has to increase the parameter $a$.

\subsubsection{The regular magnetic monopole solution: from no gravitation 
to the extremal black hole}

\begin{figure} [t]
\includegraphics*{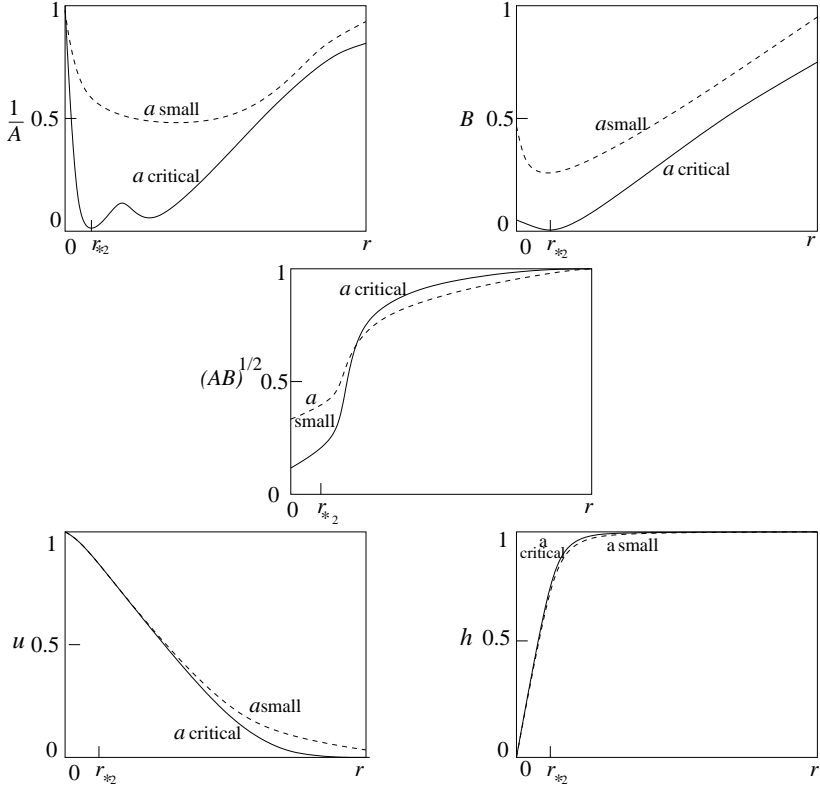}
\caption{\label{highbmonopole}
The graphs of the metric and matter functions, 
$(1/A,B,(AB)^{1/2},u,h)$, 
are plotted as a function of $r$ in the high $b$ case. The curves $a$ small 
are typical of small gravitational effects, and the curves $a$ critical 
give the properties of the extremal quasi-black hole. The radius $r_*$ is 
the the radius at which the quasi-horizon is formed, see text for details.}
\end{figure}

For low $a$ there is not much change in relation to the low $b$ case.
Low $a$ represents a highly dispersed magnetic monopole, with small
mass $M_{\rm m}$ and large radius $r_{\rm m}$. As $a$ increases the
solution gets more general relativistic and eventually gets to a black
hole, when $a=a_{\rm crit}\sim1$. An important difference to the low 
$b$ case is that instead of passing from a quasi-horizon to 
a degenerate spacetime, it passes from a quasi-horizon to
a true horizon, well inside the core at
$r_{*_2}$ \cite{lueweinberg2}.  To get a grip on the solutions we 
draw in Figure
\ref{highbmonopole} diagrams showing the metric functions and the
matter field functions as a function of $r$ for two values of $a$, $a$
small and $a_{\rm crit}$.  Specifically the behavior of the functions
is:  (i) The function $1/A$, the metric function that signals the
formation of a black hole, shows that very near $a_{\rm crit}$ there
are two radial scales, where horizon could be formed, one at $r_{*_2}$
(related to the scale set by the Higgs mass), the other at $r=r_*$
(related to the scale set by the the W mass), but at $a_{\rm
crit}$ the double zero occurs at $r_{r*_2}$, and an extremal 
horizon appears there. (ii) The metric function $B$ shows also a zero
at $r_{*_2}$ signaling the formation of an infinite redshift surface. 
Note now that $B$ is zero at one point only,  $r_{r*_2}$, 
not in a whole region as was the case for low $b$. This means 
that the configuration 
quasi-black hole with radius very near $r_{*_2}$, turns into 
a true extremal black hole rather than 
to a degenerate spacetime as in the low $b$ case.
(iii) The behavior of $(AB)^{1/2}$, which tells
whether the horizon is naked or not, confirms this behavior. It shows
that it is never zero, meaning the horizon is a regular, not a naked
one \cite{lueweinberg2}.  This means that the components of the Riemann
tensor at the horizon in an orthonormal frame are well behaved.  In
this case a particle that is sent in through the monopole, turns
around, and comes back to the point where it started, takes a proper
time $\Delta\,\tau$ which is finite, non-zero. Thus, since the Riemann
tensor is proportional to $(\Delta\,\tau)^{-2}$, as discussed in
connection to the low $b$ case, there is no funny behavior at the
horizon and all behaves well. (iv) The function $u$ for the
Yang-Mills field shows that for $a_{\rm crit}$ there is field outside
the horizon radius, i.e., there is hair. (v) The function $h$ 
for the Higgs field, behaves similarly to $u$. In the critical situation, 
it only acquires the vacuum value for radii much larger than
$r_{*_2}$.

There are three points that are worth commenting.  First, we comment
further on the behavior of $(AB)^{1/2}$ and its consequences. In terms
of the coordinate time, the particle takes for the round trip the time
$\Delta t\sim r_{*_2}\,\epsilon^{-1/2}$, where again $\epsilon\equiv
(1/A)_{\rm min}$ is a very small number. Thus the particle takes, as in
the low $b$ case, a long time to return to a coordinate observer, and
this is important in connection with entropy issues
\cite{lueweinberg2}. Indeed, in leading order, this time is determined only
from the spacetime geometry.  An observer finds that the quasi-black
hole has an inside which is inaccessible, since probes stay there for
an arbitrarily large amount of time, and describes it by a density
matrix $\rho_{\rm matrix}$ obtained by tracing over the degrees of
freedom inside the quasi-horizon, yielding an entropy $S=-\,{\rm
Tr}\left(\,\rho_{\rm matrix} \,\ln\rho_{\rm matrix}\right)$.  The
calculation of the interior entropy of a field inside a spherical box
was performed for a scalar field with the result that $S
=\gamma \frac{A}{4}$, where $\gamma$ is an undetermined
factor, and $A$ is the area of the box \cite{srednicki}. In this case,
the box is the magnetic monopole 
quasi-horizon configuration on the verge of being a
black hole.  Since one can give a little push from this configuration
to the horizon configuration, and in the latter case the entropy is
$S=\frac{A}{4}$ one can guess by continuity that the coefficient of
proportionality in the quasi-horizon case has a dependence on the size
of the box $\gamma=\gamma(r_{\rm box})$ and when a horizon forms,
$\gamma(r_{*_2})=\frac14$ \cite{lueweinberg2}.  In this sense, the
entropy of the extremal black hole is the number of the entangled degrees of
freedom inside the horizon. This analysis cannot be applied to the low
$b$ case because there is never a true horizon: in the limit, when
the object is turning into a black hole it gives a
non-smooth manifold.

Second, another feature of these monopoles is that they have a charge
to mass ratio given by $Q/M>1$. Thus if one drops neutral matter onto
a regular magnetic monopole one can form an extremal black hole
\cite{lueweinberg2}. This is contrary to the case of electric
Reissner-Nordstr\"om black holes with $Q/M<1$, and where one can drop
charged matter, with charge $q$ and mass $m$
obeying $q/m>1$, as much as one wants that one never gets
an extremal black hole (this is a version of 
the third law of black hole mechanics).

Third, one can ask what happens for $a$ higher than the critical
value.  Following \cite{lueweinberg2} one finds that there are possibly
two branches. One branch is formed of magnetically Abelian charged
Reissner-Nordstr\"om black holes. The other branch, 
has non-Abelian matter and hair outside, a horizon, and regular 
non-Abelian matter inside. Following theorems by Borde 
\cite{borde,beato1,beato2} one finds that these regular
solutions may have different inside and outside topologies. 
This issue should be further explored.

\subsubsection{The Schwarzschild black hole solution inside the monopole}

As in the low $b$ case one can put a
Schwarzschild black hole inside. This was done in \cite{brihaye1}. The
main feature is that again there is hair outside the true horizon.  The
results are in line with what we have been discussing.  Diagrams like
those of Figures 2 and 3 can be drawn, although we have not found them
in the literature.

\subsection{Further discussion}

Thus, gravitationally there are two distinct behaviors, the low $b$
case and the high $b$ case, the marginal case being at $b\simeq40$.
The low $b$ case has the following main features: 
when one turns on gravitation (when one
increases $a$) a quasi-black hole
appears from the regular monopole, which turns into a degenerate spacetime 
at the critical value $a_{\rm crit}$; 
it has a naked horizon, and shows no hair, i.e., it is
of Coulumb type, the non-trivial fields are hidden inside the horizon.
In addition, one can enrich the monopole structure by putting a
Schwarzschild black hole inside up to a certain maximum mass.  The high
$b$ case has also a regular monopole solution which, when one increases
the gravitational parameter $a$, turns into a quasi-black hole, 
and then at $a_{\rm crit}$ a true extremal
black hole appears, with regular horizon and hair. 
There is a transition between the two cases, a first order type 
transition. When $b$ is in the transition zone, there is a double 
double zero, one zero at $r_*$ and the other 
at $r_{*_2}$. So, the transition is discontinuous in radius, 
and thus in entropy. It is, however, continuous in mass \cite{lueweinberg1}. 

Other features that are also very interesting but not important in our
context are: 
(i) For very low $b$ ($b<0.1$, say), the behavior is more complicated
near $a_{\rm crit}$ \cite{breitenetal1}. If one increases $a$ from
zero, one passes $a_{\rm crit}$ up to an $a_{\rm max}$. But from
$a_{\rm crit}$ to $a_{\rm max}$ there are two solutions, one with
larger mass $M_{\rm m}$ (larger radius $r_{\rm m}$), the other with
smaller values. The one with smaller values is the one that connects
continuously with the low mass solutions. The smaller mass solutions
are stable, and so the branch which forms black holes is unstable;
(ii) For a certain range of the parameters $a$ and $b$, there are
multiple node solutions (nodes appearing in the function $u(r)$ of the
Yang-Mills field) of the type found in the Bartnick-Mckinonn solution
\cite{breitenetal1};
(iii) The particular case $b\rightarrow\infty$ in the high $b$ sector 
was analyzed in detail by Aichelburg and Bizon
\cite{bizon}. The solution has a conical singularity at $r=0$ but apart
from that it is well-behaved.  Perhaps, oddly, core behavior in
this limit was not found, we will comment on this later on. 

This program of studying the gravitational behavior of magnetic
monopoles has been continued by Brihaye et al, where the structure of
dyonic  non-Abelian black holes has been analyzed
\cite{brihaye2,brihaye3}, and gravitational monopoles in $SU(3)$,
$SU(5)$, and $SU(N)$ theories have been found
\cite{brihaye4,brihaye5,brihayeN}.

\section{Gravitational behavior of Majumdar-Papapetrou matter 
systems: two concentric spherical thin shells}

\subsection{The Majumdar-Papapetrou sector of 
the Einstein$-$Maxwell-charged$\,$dust system}

\subsubsection{The action and equations of motion}

We now want to study the Einstein-Maxwell 
system coupled to some specific 
electrically charged dust currents 
as will be described below. By dust one means a fluid with 
zero pressure. We will compare the 
configurations found  below with the magnetic configurations 
discussed in section II. A first study in this direction 
has been done in \cite{lemosweinberg} (see also 
\cite{kleberlemoszanchin}). 
The action for the Einstein$-$Maxwell-charged$\,$dust 
system is $(G=c=1)$ 
\begin{equation}
S=\int d^4x\sqrt{-g}\left(\frac{1}{16\,\pi}R+{\cal L}_{\rm matter}\right)
\,,
\label{actionEM}
\end{equation}
where $R$ is  the scalar curvature, and 
\begin{equation}
{\cal L}_{\rm matter}={\cal L}_{\rm Maxwell}+
{\cal L}_{\rm charged\,dust}+{\cal L}_{\rm int}\,.
\label{LmatterEM1}
\end{equation}
The Maxwell Lagrangian is
\begin{equation}
{\cal L}_{\rm Maxwell}=-\frac{1}{4}F_{\mu\nu}F^{\mu\nu}
\,,
\label{LmaxwellEM}
\end{equation}
\begin{equation}
F_{\mu\nu}=\partial_\mu A_\nu-\partial_\nu A_\mu
\,,
\label{F-equationEM}
\end{equation}
where $F_{\mu\nu}$ and $A_\nu$ are the electromagnetic field 
strength and potential, respectively. 
The charged dust Lagrangian, ${\cal L}_{\rm charged\,dust}$, is 
such that 
the action integral, $S_{\rm dust}=
\int d^4x \,\sqrt{-g}\,{\cal L}_{\rm dust}$, gives 
the energy-momentum tensor for charged dust, i.e., 
\begin{equation}
T_{\rm charged\,dust}^{\mu\nu}= -\frac{1}{8\pi}\,\frac{1}{\sqrt{-g}}
\frac{\delta S_{\rm charged\,dust},}{\delta\,g^{\mu\nu}}=\rho\,
u^\mu\,u^\nu\,,
\label{Ldust}
\end{equation}
where $\rho$ is the dust energy density and $u^\mu$ its 
four-velocity. The interaction Lagrangian 
${\cal L}_{\rm interaction}$ is given by 
\begin{equation}
{\cal L}_{\rm interaction}=  A_\mu\,j^\mu\,,
\label{Linteraction}
\end{equation}
where $j^\mu=\rho_{\rm e}\,u^\mu$, $\rho_{\rm e}$ being the electric
charge density.  The elementary particles are then the electromagnetic
massless photon, the massless graviton, and the massive charged dust
particles with energy density $\rho$ and charge density $\rho_{\rm
e}$. The charged dust particles may spread over a given
three-dimensional region of space, or can be
squeezed into a two-dimensional thin membrane, i.e., a shell. In the latter
case the action (\ref{actionEM}) acquires the form of a bulk action
plus a membrane action. These bulk plus membrane 
systems will be treated now.

The configuration we want to discuss is spherically symmetric, 
a star type configuration, 
with metric given again by 
\begin{equation}
ds^2=-B(r)\,dt^2+A(r)\,dr^2+
r^2\left(d\theta^2+\sin^2\theta\,d\phi^2\right)
\,,
\label{metricEMSch}
\end{equation}
and with the electric Maxwell field given by
\begin{equation}
A_0=\varphi(r)\,,\quad A_i=0\,.
\label{MfieldEMspherical}
\end{equation}
Putting this ansatz into the  Einstein$-$Maxwell-charged$\,$dust 
action (\ref{actionEM}) and 
varying the action with relation to the 
three functions, yields 
three equations, two for the gravitational 
field $B(r)$ and $A(r)$, and one for the 
Maxwell field 
$\varphi(r)$. The equations are, respectively, 
\begin{equation}
\frac{(AB)'}{AB}=8\pi\,r\,\rho\,A
\,,
\label{equationforB-EM}
\end{equation}
\begin{equation}
\left[r\left(1-\frac{1}{A}\right)\right]'
=8\pi\,r^2\,\rho+\frac{r}{A\,B}\,\varphi'^2
\,,
\label{equationforA-EM}
\end{equation}
\begin{equation}
\frac{\sqrt{B}}{r^2\sqrt{AB}}
\left[\frac{r^2}{\sqrt{AB}}\,\varphi'\right]'
=-4\pi\rho_{\rm e}
\,.
\label{equationforvarphiEM}
\end{equation}
There are three parameters in the theory: $G$ which we have set 
equal to one, $\rho$ and $\rho_{\rm e}$. Now, $\rho$ and $\rho_{\rm e}$ 
have dimensions of length to minus two. Thus one can 
form in principle two length scales. The ratio 
of these length scales yields a parameter without dimensions. 
One particular class of solutions, the one we want to treat, sets 
\begin{equation}
\frac{\rho_{\rm e}}{\rho}=1\,.
\label{majumdarcondition}
\end{equation}
(Note that the charge density 
$\rho_{\rm e}$ can have two signs, so strictly speaking one should put
$\rho_{\rm e}=\pm\rho$. In order to not carry this $\pm$ throughout we
drop the minus sign, bearing in mind that a $-$ sign can be
floating about.) Matter  obeying the condition (\ref{majumdarcondition}),  
i.e., matter with mass equal to
charge, can be called extremal charged dust in analogy with the extremal 
Reissner-Nordstr\"om black holes. 
The system of equations
(\ref{equationforB-EM})-(\ref{equationforvarphiEM}) with condition
(\ref{majumdarcondition}) is the the Majumdar-Papapetrou system
\cite{majumdar,papapetrou}.

Now, in order to show a behavior analogous to the magnetic monopole 
of the Einstein$-$Yang-Mills$-$Higgs system 
the Majumdar-Papapetrou system per se is not enough, the parameters 
do not give enough 
structure. In order to get more structure we 
have to add new parameters. First assume a given spherical 
symmetric solution, which we call a star. 
Then, a new parameter 
is the radius of the star,  $r_{\rm star}$. 
So now, one has two parameters $\rho$ and $r_{\rm star}$. 
It is preferable to swap  the star's density 
$\rho$ for the star's mass 
$M_{\rm star}$, so that the two parameters are 
$M_{\rm star}$ and $r_{\rm star}$. 
Then one can form an adimensional parameter 
\begin{equation}
a=\frac{M_{\rm star}}{r_{\rm star}}
\,.
\label{definitionofa-MP}
\end{equation}
This is the equivalent to the parameter $a$ in the 
Einstein$-$Yang-Mills$-$Higgs 
theory, see (\ref{definitionofa}). To simplify
the analysis, and without loss of generality, we can think that the  star 
is made of a thin shell of extremal charged dust, with $M_{\rm star}$ 
and $r_{\rm star}$ being now the mass and the radius of the thin shell. 
It is not difficult to see that this thin shell is a solution of 
the Majumdar-Papapetrou system \cite{kleberlemoszanchin}. 
One can now further bring into the problem a new extremal charged thin shell, 
called the secondary shell, with two new parameters, 
the mass $M_2$ and the radius $r_2$. One has then two thin shells, 
one inside the other, a configuration 
that is also a solution of the Majumdar-Papapetrou system, as will 
be displayed below. 
One can then form a 
new dimensionless parameter $b$ given by
\begin{equation}
b=\frac{M_2/r_2}{M_{\rm star}/r_{\rm star}}
\,.
\label{definitionofb-MP}
\end{equation}
This is equivalent to the secondary parameter  
of the Einstein$-$Yang-Mills$-$Higgs system appearing in 
equation (\ref{definitionofb}). 
This double shell solution has 
four parameters 
$M_{\rm star},r_{\rm star},M_2,r_2$. In order to 
produce the required model one  
should restrict these four parameters through a 
constraint equation, as in 
the magnetic monopole case. Generically, the two shells are 
indistinguishable, 
one cannot say whether the outer one is the star or the 
secondary shell. To be definitive, the inner shell it is called   
the secondary shelll, the outer shell is the star, and 
we keep the secondary shell always inside the star, through 
the constraint  
\begin{equation}
r_{\rm star}=2\,r_2\,.
\label{constraint-MP}
\end{equation}
The factor $2$ in (\ref{constraint-MP}) was chosen for convenience, 
any real number greater than one will do. 
Equation (\ref{constraint-MP}) is the equivalent to 
the constraint (\ref{constraint1}) in the magnetic monopole case.  
Thus the system we are going to work with is a Majumdar-Papapetrou
system with two extremal charged shells.
This simple system mimics a good deal of behavior of the
Einstein$-$Yang-Mills$-$Higgs system. Instead 
of working with thin shells, one could work with the 
thick shell solutions found in \cite{kleberlemoszanchin} or with 
the Bonnor stars \cite{bonnor2,bonnor3,bonnor4}, but this 
only complicates the technical analysis of the 
problem without further illuminating it.

\subsubsection{Some properties and scales of 
the Majumdar-Papapetrou double shell}

\begin{figure} [t]
\includegraphics*{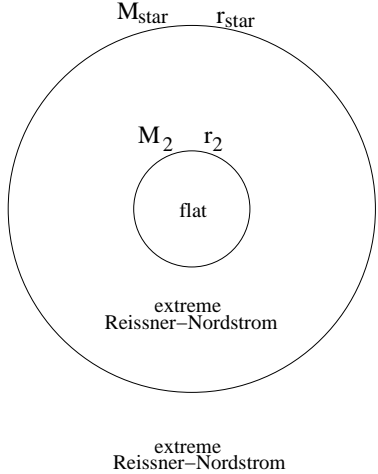}
\caption{\label{shellwithinshell}
A schematic drawing of the double shell solution in 
the Majumdar-Papapetrou system, showing 
the secondary shell inside the star shell.} 
\end{figure}

We are now ready to put a shell within a shell, and simulate the
behavior of the gravitational magnetic monopoles. 
The star (outer shell) and the 
secondary shell (inner shell) are considered to be infinitesimally thin, 
see Figure \ref{shellwithinshell}.
Then, the  metric valid from 
$0\leq r<\infty$, for a Majumdar-Papapetrou 
spacetime with two extremal matter thin shells, is given by 
\begin{equation}
ds^2= -\frac{    \left(1-\frac{M}{r_{\rm star}}\right)^2 
\left(1-\frac{M_2}{r_2}\right)^2}
{\left(1-\frac{M_2}{r_{\rm star}}\right)^2}\,dt^2
+dr^2+r^2\left( d\theta^2+\sin^2\theta\,d\phi^2\right)
\,,\quad 0\leq r\leq r_2
\,, 
\label{metrictwoshellinteriorflat}
\end{equation}
\begin{equation}
ds^2= -\frac{ \left(1-\frac{M}{r_{\rm star}}\right)^2}
{\left(1-\frac{M_2}{r_{\rm star}}\right)^2}
\left(1-\frac{M_2}{r}\right)^2\,dt^2
+\frac{dr^2}{\left(1-\frac{M_2}{r}\right)^2}+
r^2\left( d\theta^2+\sin^2\theta\,d\phi^2\right)
\,,\quad r_2\leq r\leq r_{\rm star}
\,, 
\label{metrictwoshellinbetween}
\end{equation}
\begin{equation}
ds^2= -
\left(1-\frac{M}{r}\right)^2
\,dt^2
+\frac{dr^2}{\left(1-\frac{M}{r}\right)^2}+
r^2\left( d\theta^2+\sin^2\theta\,d\phi^2\right)
\,,\quad r_{\rm star}\leq r<\infty
\,, 
\label{metrictwoshellexterior}
\end{equation}
where $M=M_{\rm star}+M_2$ is the total mass of the system. 
The constants in the $g_{tt}$ components were chosen so that the
metric matches at the shells.
The electric field is 
\begin{equation}
\varphi=1-
\frac{    \left(1-\frac{M}{r_{\rm star}}\right) 
\left(1-\frac{M_2}{r_2}\right)}
{\left(1-\frac{M_2}{r_{\rm star}}\right)}
\,,\quad 0\leq r\leq r_2
\,, 
\label{electricfieldtwoshellinteriorflat}
\end{equation}
\begin{equation}
\varphi=1-
\frac{ \left(1-\frac{M}{r_{\rm star}}\right)\left(1-\frac{M_2}{r}\right)}
{\left(1-\frac{M_2}{r_{\rm star}}\right)}
\,,\quad r_2\leq r\leq r_{\rm star}
\,, 
\label{electricfieldwoshellinbetween}
\end{equation}
\begin{equation}
\varphi=1-
\left(1-\frac{M}{r}\right)=\frac{M}{r}
\,,\quad r_{\rm star}\leq r<\infty
\,. 
\label{electricfieldtwoshellexterior}
\end{equation}
The fluid field is given by the surface energy densities of the shells. 
For the secondary thin shell one has that the surface energy density 
$\sigma_2$ is given by 
\begin{equation}
\sigma_2=\frac{M_2}{4\pi r_2^2}\,,
\label{surfacedensitysecondary}
\end{equation}
with the the corresponding surface electric charge density of the shell 
$\sigma_{{\rm e}_2}$ given by 
$\sigma_{{\rm e}_2}=\sigma_2$.
For the thin shell star one has that the surface energy density 
$\sigma_{\rm star}$ is given by 
\begin{equation}
\sigma_{\rm star}=\frac{M_{\rm star}}{4\pi r_{\rm star}^2}\,,
\label{surfacedensitystar}
\end{equation}
with the corresponding surface electric charge density of the shell 
$\sigma_{{\rm e}_{\rm star}}$ 
given by $\sigma_{{\rm e}_{\rm star}}=\sigma_{\rm star}$. 
Note that the $g_{rr}$ component 
of the metric has a step function at $r_2$ and $r_{\rm star}$. 
This is no problem, one can smooth it out by considering a 
shell with small thickness \cite{kleberlemoszanchin}, 
but for the problem we are considering it is irrelevant. 

One important question is which shell, and in which conditions 
a shell, forms a horizon. 
We know that a horizon should form when $1/A=0$. Suppose 
a $b$ is given, and one starts to increase 
$a$. Then it is meaningfull to ask which shell forms first a horizon, 
the star or the secondary shell? 
To answer it note that
\begin{equation}
\left.\frac{1}{A}\right)_{r_2}=1-\frac{M_2}{r_2}=1-ab
\,,
\label{themetricfunctionAatinteriorshell}
\end{equation}
\begin{equation}
\left.\frac{1}{A}\right)_{r_{\rm star}}
=1-\frac{M}{r_{\rm start}}=1-\frac{M_{\rm star}}{r_{\rm star}}
\left(
1+\frac{M_2/r_2}{M_{\rm star}/r_{\rm star}}\frac{r_2}{r_{\rm star}}
\right)
=1-a\left(1+\frac{b}{2}\right)
\,. 
\label{themetricfunctionAatexteriorshell}
\end{equation}
\centerline{}
It is then clear that there are three cases:  
\begin{description}
\item[$b<2$ $-$ ] when $a$ increases an external horizon forms 
first at $r_{\rm star}=M$, with  $a_{\rm crit}=1/(1+b/2)$. This is 
analogous to the behavior of magnetic monopoles with low $b$, 
where an external horizon forms outside the core. 
\item[$b>2$ $-$ ] when $a$ increases an interior horizon forms 
first at $r_2=M_2$, with $a_{\rm crit}=\frac{1}{b}$. 
This is 
analogous to behavior of magnetic monopoles with high $b$, 
where an external horizon forms within the core. 
\item[$b=2$ $-$] when $a$ increases a horizon forms at both 
shells, interior and exterior
with $a_{\rm crit}=1/2$. This divides the two cases above. 
\end{description}
As we will show below the solution  does
not develop a true horizon. Independently of $b$, upon increasing $a$, 
a quasi-horizon appears. Then at the critical value one gets a degenerated 
spacetime, and for values of $a$ above the critical there is 
no static solution, the shell collapses (see \cite{gaolemos})
into a singularity and an extremal Reissner-Norstr\"om black hole 
forms.
In what follows we study each type of configuration. We
start with low $b$, $b<2$, then we do high $b$, $b>2$.

\subsection{The gravitational behavior as a function of $a$  
for low $b$ }

Since $b=\frac{M_2/r_2}{M_{\rm star}/r_{\rm star}}$, low $b$ can be seen as 
a relatively small secondary mass $M_2$, or large secondary
radius $r_2$, which means that the secondary shell has little
influence in the structure.  Given a low $b$ configuration, we want to
understand how the structure changes as the parameter 
$a=\frac{M_{\rm star}}{r_{\rm star}}$ increases.
We present plots giving the behavior of the metric 
and matter functions as a function of radius for typical cases, 
discuss the naked
horizon behavior and the Coulumb character 
of these solutions, put an extremal Reissner-Nordstr\"om 
black hole inside, and sketch some diagrams covering the space of
solutions.

\subsubsection{The regular Majumdar-Papapetrou 
double shell solution: from no gravitation 
to the extremal quasi-black hole and beyond}

\begin{figure} [t]
\includegraphics*{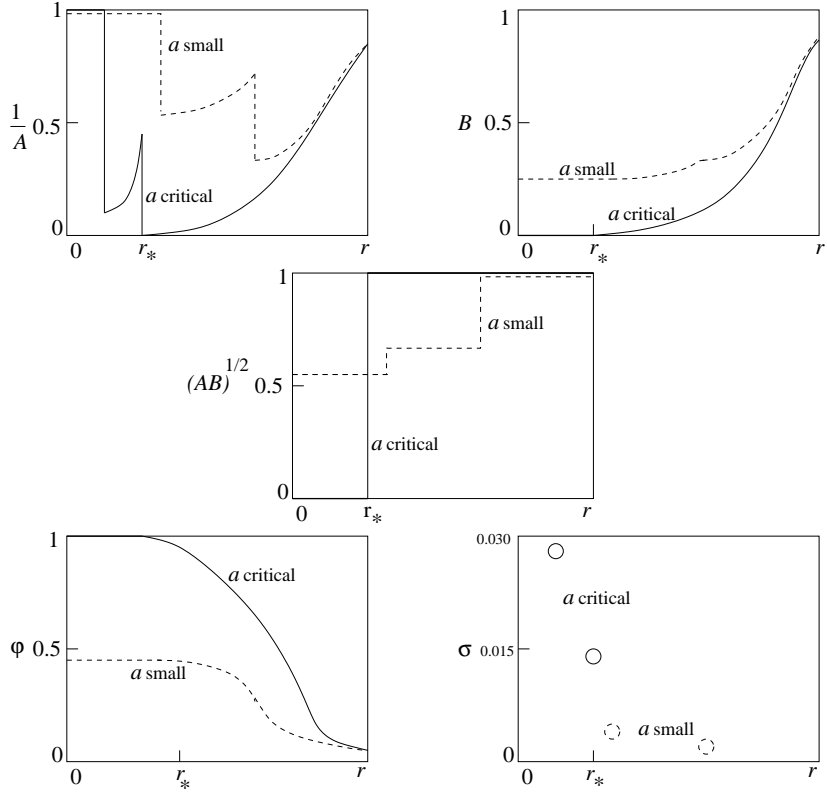}
\caption{\label{MPlowbstar}
The graphs of the metric and matter functions, 
$(1/A,B,(AB)^{1/2},\varphi,\sigma)$, 
are plotted as a function of $r$ in the low $b$ ($b<2$) case. 
The curves $a$ small 
(dashed lines) 
are typical of small gravitational effects, and the curves $a$ critical 
(full lines) 
give the properties of the extremal electrical quasi-black hole. 
The radius $r_*$ is $r_{\rm star}$ at the quasi-horizon (arbitrarily near 
the critical radius),  
see text for details.
(In the graphs, we have used $b=1$ as a typical low $b$ case, and 
have set $M_{\rm star}=\frac{5}{2}$, $M_2=\frac{5}{4}$, 
and for $a$ small we have put $r_{\rm star}=10$, $r_2=5$, 
while for $a$ critical  we have put 
$r_{\rm star}=\frac{15}{4}$, $r_2=\frac{15}{8}$.)}
\end{figure}

We have seen from Equations
(\ref{themetricfunctionAatinteriorshell})-(\ref{themetricfunctionAatexteriorshell})
that for fixed $b$, with $b<2$, an extremal quasi-horizon appears 
when the parameter $a$ increases, i.e., when one puts more
gravitation into the star shell. 
A small $a$ parameter, i.e., $M_{\rm star}/r_{\rm star}$ small, 
means that the star 
shell is very dispersed. As $a$ increases, eventually it gets to a 
stage where a kind of an extremal event horizon forms. 
Using Equations 
(\ref{metrictwoshellinteriorflat})--(\ref{themetricfunctionAatexteriorshell}) 
one can plot the behavior of 
the metric and matter functions as a function of radius for two 
values of $a$, $a$ small, and $a$ arbitrarily near 
$a_{\rm crit}$, when a quasi-horizon forms, 
see Figure \ref{MPlowbstar}. 
Specifically, the behavior of the functions  $1/A$, $B$, $(AB)^{1/2}$, 
$\varphi$, and $\sigma$ is: 
(i) The function $1/A$ signals the formation of a black hole.  For $a$
small the function $1/A$ starts at the value 1 (thus there are no
conical singularities) drops slightly at the secondary 
shell, rises and drops again at the star shell, and
then rises again to 1 at infinity. When $a=a_{\rm crit}$ 
(or arbitrarily near it) the function
gets a `double' zero at $r=r_*$ (it would be a double
zero had we smoothed out enough the matter) signaling the formation
of a kind of an extremal Reissner-Nordstr\"om black hole. 
(ii) The function $B$, the redshift function, has the usual behavior for 
$a$ small. However, for $a_{\rm crit}$ the whole of the region 
inside $r_*$ gets infinitely redshifted. This means 
that the manifold is not smooth. Thus the critical case is not 
a true black hole, it is a degenerated spacetime.
(iii) The product function $(AB)^{1/2}$ is important to determine
whether the forming horizon is naked or not.  We find
that a particle on a return trip to the star takes a proper time given
by $\Delta \tau\sim r_*\,\epsilon^{1/2}$, where $r_*$ is the radius of the
quasi-horizon, and, near the critical solution, 
$\epsilon\equiv (1/A)_{\rm min}$
is a very small quantity \cite{kleberlemoszanchin}.  Since this proper time
is arbitrarily small, the Riemann tensor diverges at the horizon, and
the horizon is naked. For completeness we give the coordinate time 
$\Delta t$ taken by the particle in its trip, 
$\Delta t\sim r_*\,\epsilon^{-1/2}$, implying that 
the particle takes a long time
to return for a coordinate observer.
(iv) The function $\varphi$ tells whether the solution 
has hair or not. It  starts constant, then decays with 
$1/r$, with a bump at $r_2$ and at $r_{\rm star}$. When the 
horizon forms the field is a pure Coulumb field, showing no-hair. 
(v) The surface density function $\sigma$ of the charged dust is also 
drawn, for completeness. Outside the quasi-horizon at $r_*$ 
there is no matter.

The case $b=0$ is worth discussing because it is the simplest one in
the low $b$ sector. There is no secondary shell ($M_2=0$) and so it
represents a single thin shell with mass $M_{\rm star}$ and radius
$r_{\rm star}$. It is interesting because on one hand it has the same
properties of any other low $b$ case, on the other hand, it is easier
to figure out what happens above criticality, i.e., for $a>a_{\rm
crit}$ ($M_{\rm star}>r_{\rm star}$).  We have seen that when the
precise equality holds, $a= a_{\rm crit}$, the redshift function $B$
is zero not only at the horizon but also in the whole region inside,
meaning that in fact a true black hole does not form, since inside
there is no smooth manifold. For $a>a_{\rm crit}$ one seems to have
now a shell of matter at $r_{\rm star}$ inside an extremal electrical
Reissner-Nordstr\"om black hole at $r_{\rm bh}=M_{\rm star}$, the
solution being everywhere free from curvature singularities.  
Following a theorem by Borde \cite{borde}, this
would mean that the topology of spacelike slices in this black hole
spacetime would change from a region where they are noncompact to a
region where they are compact, in the interior.  In our case this in
fact does not happen, there are no solutions with $a>a_{\rm crit}$, 
i.e., $m>r_{\rm o}$, the
shell collapses into a singularity \cite{gaolemos}.

\subsubsection{Non-regular Majumdar-Papapetrou shell solutions: The
extremal Reissner-Nordstr\"om black hole solution inside the thin
shell star} 

One can put a black hole inside the double thin shell and obtain a
structure similar to the one found when one puts a black hole inside a
magnetic monopole. For the double thin shell, the extra inner black
hole has to be an extremal Reissner-Nordstr\"om black hole, rather
than a Schwarzschild black hole, to keep the solutions within the
Majumdar-Papapetrou system.  If one puts a non-extremal black hole
foreign tensions would develop at the thin shells. So, in order to
stick to pure Majumdar-Papapetrou system we stick to an inner extremal
Reissner-Nordstr\"om black hole.  In order to simplify the analysis,
we will work with the $b=0$ which is a good simple case for low $b$.
For any other small $b$, such that $b<2$, the result is analogous.  In
the $b=0$ case one has $M_2=0$. Thus the system is formed by the star
shell with mass $M_{\rm star}$ and radius $r_{\rm star}$, and an inner
extremal Reissner-Nordstr\"om black hole with mass $M_{\rm bh}$ and
radius $r_{\rm bh}$ ($M_{\rm bh}=r_{\rm bh}$). 
The metric is now
\begin{eqnarray}
ds^2&=& -\left(1-\frac{M_{\rm bh}}{r}\right)^2 
\frac{\left(1-\frac{M}{r_{\rm star}}\right)^2}
{\left(1-\frac{M_{\rm bh}}{r_{\rm star}}\right)^2}
\,dt^2
+\frac{dr^2}{\left( 1-\frac{M_{\rm bh}}{r}\right)^2}+
\nonumber
\\
&&r^2\left( d\theta^2+\sin^2\theta\,d\phi^2\right)
\,,\quad 
0\leq r\leq r_{\rm star}
\,, 
\label{metrictblackholeshell1}
\end{eqnarray}
\begin{equation}
ds^2= -\left(1-\frac{M}{r}\right)^2 
\,dt^2
+\frac{dr^2}{\left( 1-\frac{M}{r}\right)^2}+
r^2\left( d\theta^2+\sin^2\theta\,d\phi^2\right)
\,,\quad r_{\rm star}\leq r
\,,
\label{metrictblackholeshell2}
\end{equation}
where here $M=M_{\rm star}+ M_{\rm bh}$ is now the total mass. 
The electric field $\varphi(r)$ and the charge density field 
$\rho(r)=\sigma_{\rm star}(r_{\rm star})$ 
profile accordingly.

\begin{figure} [t]
\includegraphics*[height=2.5in]{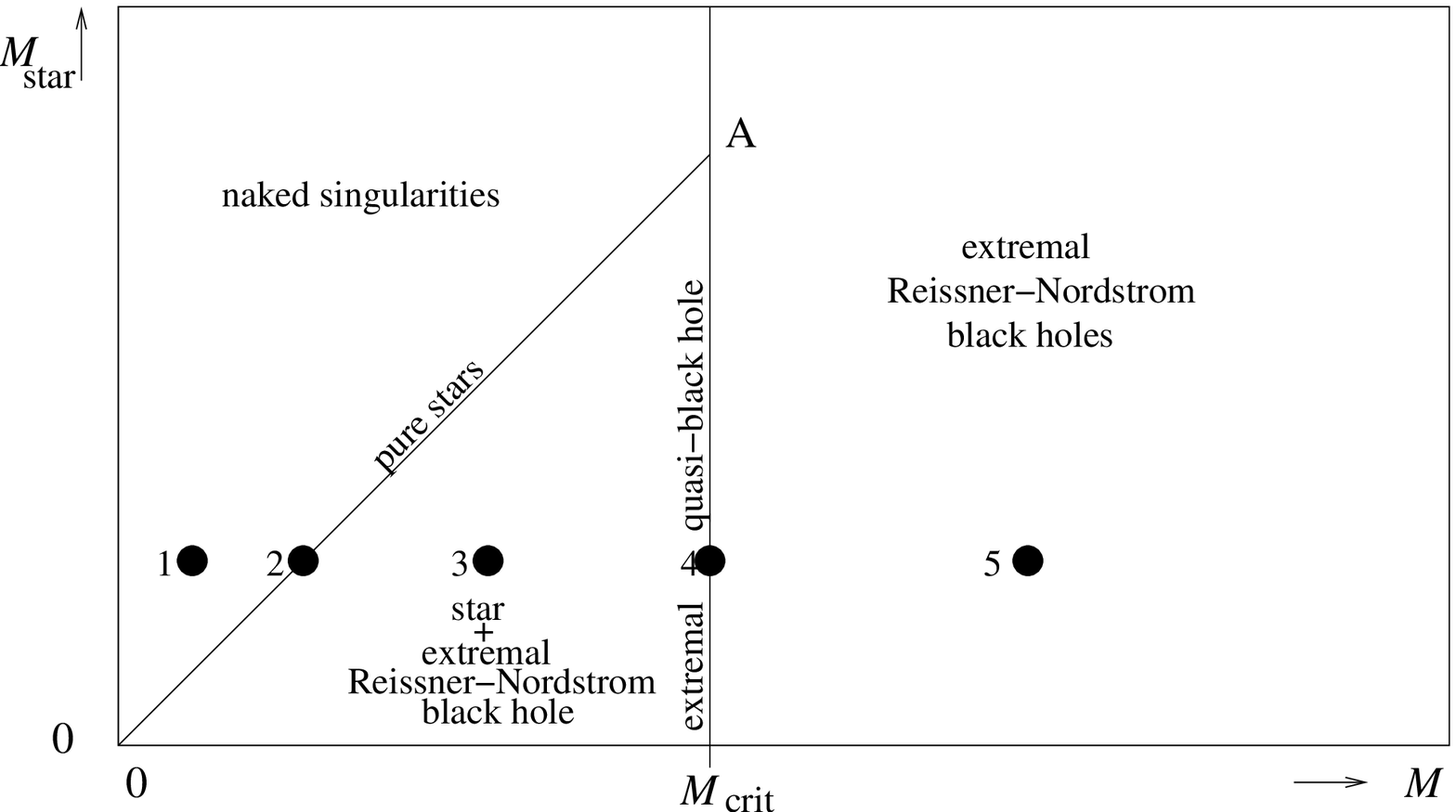}
\vskip 5mm
\includegraphics*[height=1.1in]{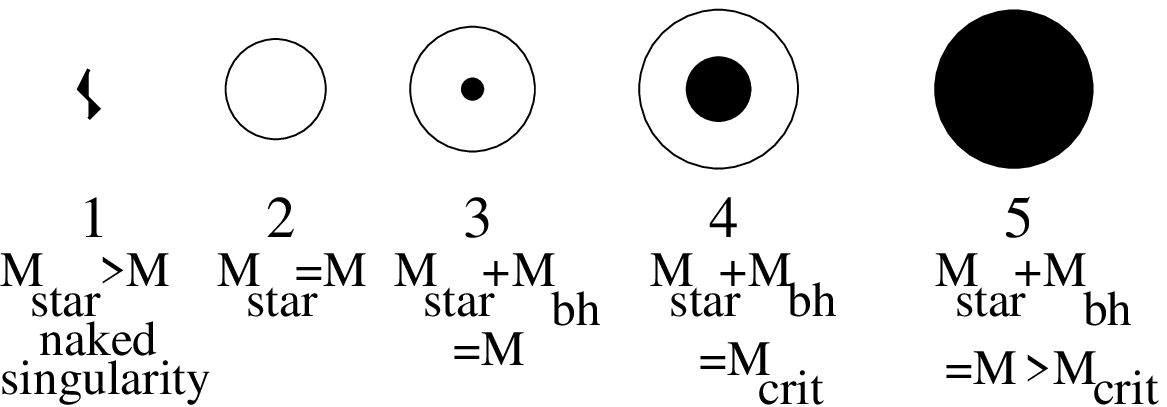}
\caption{\label{graphMPstarmassversustotalmass}
The space of solutions in a $M_{\rm star}\times M$, where 
$M=M_{\rm star}+M_{\rm bh}$ is the total mass, is plotted.
For each point 1-5, along a constant star mass, 
in the diagram, the corresponding 
configuration is pictorially represented in the bottom part 
of the figure.
This is the graph made in Figure  \ref{graphmonopolemassversustotalmass}
(see also \cite{leenairweinbergPRD92}) 
adapted to the thin shell Majumdar-Papapetrou system.}
\end{figure}
\begin{figure} [t]
\includegraphics*[height=2.5in]{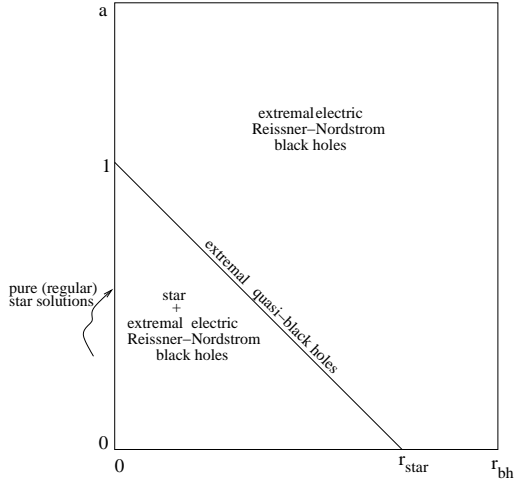}
\caption{\label{aMPversusrblackhole}
The space of solutions in an $a\times r_{\rm bh}$ plot, where
$a=\frac{M_{\rm star}}{r_{\rm star}}$, and $r_{\rm bh}$ is the radius
of the Reissner-Nordstr\"om black hole inside the star.  This is the
graph drawn in Figure \ref{aversusrblackhole} (see also
\cite{breitenetal1}) adapted to the thin shell Majumdar-Papapetrou
system.}
\end{figure}

To understand the generic behavior of this system it is helpful to
make a plot of the solution space, similar to the 
plot made for a Schwarzschild black hole inside the magnetic monopole 
shown in Figure
\ref{graphmonopolemassversustotalmass}.  We do this in Figure
\ref{graphMPstarmassversustotalmass}, where we plot the solution space
in a $M_{\rm star}\times M$ diagram for fixed $r_{\rm star}$. 
There are three regions and
two lines. The pure star line, i.e., the regular solutions discussed
above with the total mass equal to the star mass, is represented by a
line with slope 1. The top-left region represents extremal charged naked
singularities.  The center-left region represents star+(extremal
Reissner-Nordstr\"om black hole) solutions displayed in Equations
(\ref{metrictblackholeshell1})-(\ref{metrictblackholeshell2}).  At 
values arbitrarily near the
critical mass $M_{\rm crit}$ the solutions are extremal electric
charged Reissner-Nordstr\"om quasi-black holes, which 
degenerate at the critical value.  To the right there is
a region of a totally collapsed 
shell star inside an extremal Reissner-Nordstr\"om black
hole, which means the solutions represent pure
extremal Reissner-Nordstr\"om black
holes. We show pictorially each representative configuration along a
constant star mass line. Each numbered point (from 1 to 5) in
Figure \ref{graphMPstarmassversustotalmass} is represented in the
bottom of the figure by a schematic drawing.  We see that taking
$M_{\rm star}$ constant and increasing $M$ we pass through point 1
where $M_{\rm star}$ is greater than $M$ and therefore there is a negative
mass at the center, through point 2 where one finds a thin shell
solution with $M_{\rm star}=M$, through point 3 where there is a black
hole inside the star, through point 4 which is the case arbitrarily 
near the critical value where
$M_{\rm star}+M_{\rm bh}=r_{\rm star}$, and thus an extremal
quasi-black hole appears at $r_{\rm star}$, finally to point 5 where
$r_{\rm star}$ has collapsed inside the horizon radius 
to form an extremal black
hole. Note there is a jump in horizon radius from a point
infinitesimally to the left of point 4, to a point
infinitesimally to the right of point 4. 
Note also that this diagram is done for fixed $r_{\rm star}$. 
For another value of $r_{\rm star}$, one gets the
same diagram, but with the vertical critical line critical
shifted, to the right when the new $r_{\rm star}$ is larger, 
and to the left when  the new $r_{\rm star}$ is smaller than the 
original value. Comparison of the Figures 
\ref{graphmonopolemassversustotalmass} and
\ref{graphMPstarmassversustotalmass} shows the similarities between
the magnetic monopole and the Majumdar-Papapetrou system.

One can also translate Figure \ref{aversusrblackhole} into this
Majumdar-Papapetrou system. This is done in Figure
\ref{aMPversusrblackhole}, where we display the important regions in a
graph $a\times r_{\rm bh}$, where again 
$a=\frac{M_{\rm star}}{r_{\rm star}}$, 
and $r_{\rm bh}$ is the radius of the Reissner-Nordstr\"om 
black hole inside the star. 
There are two regions and two lines.  There is the vertical
line, $r_{\rm bh}=0$, of regular star solutions.  There is the region
where star+(extremal electric Reissner-Nordstr\"om black hole)
solutions exist.  There is the line where the system forms a
quasi-black hole (i.e., a solution arbitrarily near the critical
degenerate case). Finally there is the region where an extremal
electric black hole exists. The naked singularity region, not shown,
would appear for negative $r_{\rm bh}$, i.e., for negative 
black hole masses, $r_{\rm bh}=M_{\rm bh}<0$.

\subsection{The gravitational behavior as a function of $a$ 
for high $b$}

High $b$ can be seen as 
a relatively large secondary mass $M_2$, or small secondary
radius $r_2$, which means that the secondary shell has  a decisive 
influence in the structure.  Given a high $b$ configuration, we want to
understand how the structure changes as the parameter 
$a=\frac{M_{\rm star}}{r_{\rm star}}$ increases.
We present plots giving the behavior of the metric 
and matter functions as a function of radius for typical cases, 
discuss the naked
horizon behavior and the non-Coulumb character 
of these solutions, and we briefly
comment on putting an extremal Reissner-Nordstr\"om 
black hole inside the high $b$ double shell system.

\subsubsection{The regular solution: from no gravitation 
to the extremal quasi-black hole and beyond}

\begin{figure} [t]
\includegraphics*{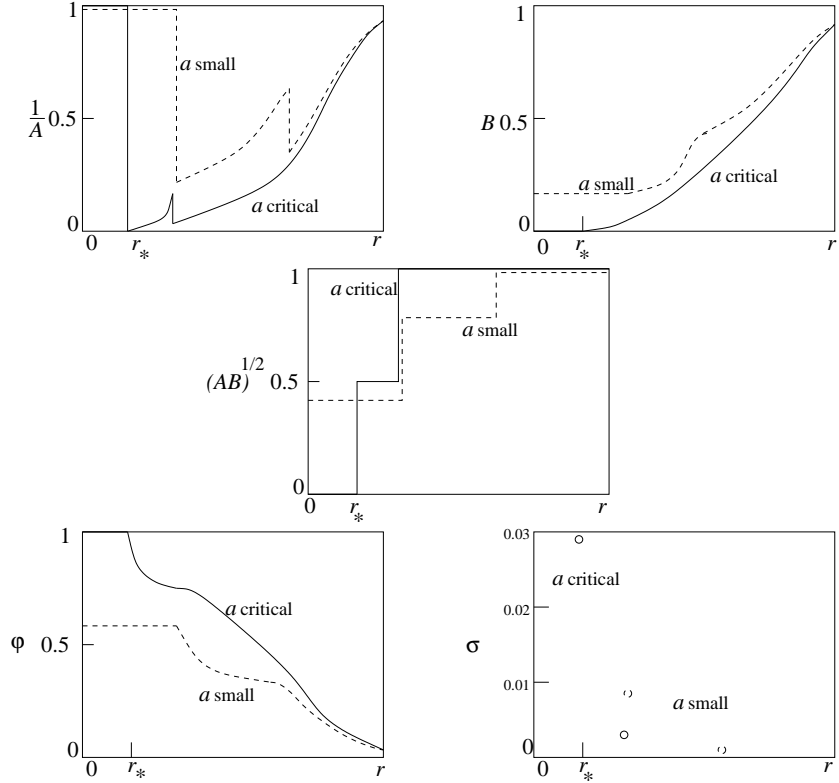}
\caption{\label{MPhighbstar}
The graphs of the metric and matter functions, 
$(1/A,B,(AB)^{1/2},\varphi,\sigma)$, 
are plotted as a function of $r$ in the high $b$ ($b>2$) case. 
The curves $a$ small 
(dashed lines) 
are typical of small gravitational effects, and the curves $a$ critical 
(full lines) 
give the properties of the extremal quasi-black hole. 
The radius $r_*$ is now $r_2$ at the quasi-horizon, 
see text for details.
(In the graphs, we have used $b=4$ as a typical high $b$ case, and 
have set $M_{\rm star}=\frac{5}{4}$, $M_2=\frac{5}{2}$, 
and for $a$ small we have put $r_{\rm star}=10$, $r_2=5$, 
while for $a$ critical  we have put 
$r_{\rm star}=5$, $r_2=\frac{5}{2}$.)}
\end{figure}

In contrast with low $b$, in the high $b$ case, 
an extremal quasi-horizon forms at 
the secondary shell $r_2$, rather than in the star shell. 
Using Equations 
(\ref{metrictwoshellinteriorflat})--(\ref{themetricfunctionAatexteriorshell}) 
one can draw  the
important field functions as a function of $r$, for a given high $b$ 
and for two values of $a$, $a$ small and $a$ critical, 
see Figure \ref{MPhighbstar}. 
The behavior of the functions  $1/A$, $B$, $(AB)^{1/2}$, 
$\varphi$, and $\sigma$ is: 
(i) For $a$
small the function $1/A$ starts at the value 1 
drops at the secondary 
shell, rises and drops slightly at the star shell, and
then rises again to 1 at infinity.
When $a$ is arbitrarily near $a_{\rm crit}$ the function
gets a double zero at $r=r_*$, now situated at 
the secondary shell $r_2$, 
signaling the formation of a quasi-horizon. 
(ii) For $a$ small, 
the function $B$ has the usual behavior. 
For $a$ arbitrarily near $a_{\rm crit}$, $B$ is 
arbitrarily near zero throughout the region  
inside $r_2$. At $a=a_{\rm crit}$ precisely the manifold is 
not smooth. Thus again, the critical case is not 
a true black hole, it is a degenerate manifold. 
(iii) The function $(AB)^{1/2}$ 
is zero inside the quasi-horizon confirming the existence of a naked 
behavior. This is not quite 
the same 
as the high $b$ behavior for the magnetic monopole, since 
the magnetic system gets in the high $b$ case a non-naked horizon. 
(iv) At $a$ arbitrarily near $a_{\rm crit}$, the function $\varphi$ 
does not have a Coulumb type behavior as one can see from Figure 
\ref{MPhighbstar}, the elctric field outside the quasi-horizon gets a bump 
due to the presence of the outer shell. Strictly speaking one cannot talk 
of a no-hair violation since the no-hair theorem is applied to black 
holes, not quasi-black holes.  
(v) The surface density function $\sigma$ of the charged dust is also 
drawn, for completeness. 

There are two questions that can be asked. The first one is what
happens if one increases $a$ past $a_{\rm crit}$. For $a>a_{\rm crit}$
one gets an extremal electric Reissner-Nordstr\"om black hole outside
the shell with radius $r_2$.
This secondary shell then collapses, leaving an extremal black 
holes with a star shell outside. This is then analogous 
to the extremal black hole solution inside
the star shell discussed previously in the low $b$ case. Upon
increasing $a$ further one hits a new critical value, $a_{\rm
crit\,new}$, where a new horizon forms at the exterior star shell
$r_{\rm star}$.

The second question is what happens when $b\rightarrow\infty$.  In the
magnetic monopole system this case has been analyzed in \cite{bizon}.
When one increases $b$, keeping $a$ fixed, one finds that $r_2$ gets
relatively smaller and smaller. The behavior is best displayed by
looking into the $1/A$ plot of Figure \ref{MPhighbstar}. For $a$ small
and fixed, when one increases $b$ the minimum at $r_2$ is displaced
more and more toward $r=0$. Eventually at $b\rightarrow\infty$ the
minimum hits the $r=0$ line at a point $1/A$ less than one, which
means that the configuration starts at a conical singularity. This
example shows why \cite{bizon} did not get the high $b$ behavior
found in \cite{lueweinberg1}, namely, 
a smooth black hole formation in the core of the magnetic monopole. 
What happens is that for
$b\rightarrow\infty$ a horizon (a kind of singularity) in the inner
secondary shell does not form at the core because the initial
configuration already possesses at the core ($r=0$) a conical
singularity (another kind of singularity which substitutes the horizon
in this limit $b\rightarrow\infty$). This conical configuration exists
for a given typical value of $a$.  Upon increasing $a$ further one
hits a critical value for $a$ (corresponding to the $a_{\rm
crit\,new}$ mentioned above) where a new quasi-horizon forms at the
exterior star shell $r_{\rm star}$.

\subsubsection{The extremal black hole solution inside the system}

As in the low $b$ case, where an extremal black hole was put inside 
the low $b$ shells, one can also put an extremal black hole 
inside the high $b$ shells. We will not do this
here since the behavior is similar to the previous cases.  
In the magnetic monopole high $b$ case this was done in 
\cite{brihaye1} .

\subsection{Further discussion}

{\it (i) The $b=2$ configuration:} 
\vskip 0.3cm

We have treated the cases $b<2$ and $b>2$. The case  $b=2$ 
is also worth commenting as a limiting case. 
The new feature is that at $a_{\rm
critical}$ the function $1/A$ develops two double zeros, one at 
the secondary shell $r_2$, the other at the 
star shell $r_{\rm star}$. Thus, on going from
$b<2$ to $b>2$ the quasi-horizon jumps discontinuously in radius at
some critical $a$. If the entropy of this object can be related to the
area of the object, as was done in \cite{lueweinberg2}, then the
entropy also jumps discontinuously when one passes from $b<2$ to
$b>2$.  In the transition there is no mass jump, the mass is
continuous, so that it is a kind of first order phase transition.

\vskip 0.5cm
{\it (ii) More complex configurations:}
\vskip 0.3cm

One can put a third extremal matter shell inside the other two. 
In this case one has two new parameters, $M_3$ and $r_3$, and a 
new dimensionless parameter 
$c$ can be given. In analogy with $a$ and $b$ of 
equations (\ref{definitionofa-MP}) and (\ref{definitionofb-MP}), 
one finds, 
\begin{equation}
c=\frac{M_3/r_3}{M_2/r_2}
\,.
\label{definitionofc-MP}
\end{equation}
Assume also as the constraint equation that $r_2=2\,r_3$. Then, one has 
\begin{eqnarray}
\left.\frac{1}{A}\right)_{r_{\rm star}}=
            1-\frac{M_{123}}{r_{\rm star}}=
            1-a\left[1+\frac{b}{2}\left(1+\frac{c}{2}\right)\right]\,,\\
\left.\frac{1}{A}\right)_{r_2}=1-\frac{M_{23}}{r_2}=
                                1-ab\left(1+\frac{c}{2}\right)\,,\\
\left.\frac{1}{A}\right)_{r_3}=1-\frac{M_3}{r_3}=1-abc
\,,
\label{athemetricfunctionAatexteriorshell}
\end{eqnarray}
where $M_{123}=M_{\rm star}+M_2+M_3$ is the total mass of the system,  
and $M_{23}=M_2+M_3$. 
Then one has: (I) For $b<\frac{4}{2+c}$ and $b<\frac{4}{3c-2}$ 
a quasi-horizon forms first at $r_{\rm star}$, 
with $a_{\rm crit}=1/[1+b/2(1+c/2)]$.
(II) For $b>\frac{4}{2+c}$ and $c<2$ a 
quasi-horizon forms first at $r_2$, 
with $a_{\rm crit}=1/b(1+c/2)$.
(III)  For the two cases (i) $b<\frac{4}{2+c}$ and $b>\frac{4}{3c-2}$, 
and (ii) $b>\frac{4}{2+c}$ and $c>2$, 
a quasi-horizon forms first at $r_3$ 
with $a_{\rm crit}=1/(bc)$. Equalities mean that 
the three quasi-horizons form together 
with $b=1$, $c=2$ and  $a_{\rm crit}=1/2$. 
Two quasi-horizons alone cannot form together. 

One can continue to put more shells with 
the emergence of ever more complex 
behavior in the function $1/A$. This type of behavior 
should also happen in non-Abelian theories  
with more Higgs scales.

\vskip 0.5cm
{\it (iii) Other configurations:}
\vskip 0.3cm
Other configurations that could be dealt with 
are a thick shell within a thin shell, with the thick shell 
being the solution found in \cite{kleberlemoszanchin}.
The behavior is similar to what we have been discussing. 
For the low $b$ case it will give for $a_{\rm crit}$ an extremal
naked, Coulumb (no-hair), quasi-black hole. For high $b$ it would give an 
extremal naked, non-Coulumb 
(hair), quasi-black hole.  One can also put an extremal black hole
inside a thick shell, although there is no known exact solution for
it.

\section{Conclusions}

We have shown that gravitational magnetic monopoles and
Majumdar-Papapetrou stars, in the form of two thin shells, have common
properties.  We have shown that both systems have extremal quasi-black
hole solutions, some without hair while others developing some type of
hair.  Both, the monopole system and the two shell Majumdar-Papapetrou
system, possess solutions with naked behavior, i.e., tidal forces tend
to infinity at the quasi-horizon.  At the critical value the interior
solution does not give a smooth manifold. 
For other parameters in the space of solutions of the
magnetic monopole system, specifically for high Higgs mass, there are
solutions with non-naked behavior, allowing the formation of a true
black hole.  On the other hand, the two shell Majumdar-Papapetrou
system, never shows non-naked behavior, there are only quasi-black
hole solutions.  In both systems one can put a black hole inside the
configuration without destabilizing the system, for a range of
parameters.

\acknowledgments 

JPSL acknowledges financial support from
the Portuguese Science Foundation FCT and FSE, through
POCTI along the III Quadro Comunitario de Apoio, reference number
SFRH/BSAB/327/2002 and through project PDCT/FP/50202/2003, 
and thanks Columbia University for hospitality.
VTZ thanks Observat\'orio Nacional-Rio de Janeiro for hospitality.


\end{document}